\documentclass[structabstract]{aa}  
\usepackage{graphicx}
\usepackage{txfonts}
\usepackage{natbib}
\bibpunct{(}{)}{;}{a}{}{,}
\usepackage{longtable,lscape}
\usepackage{appendix}

\def\Tiny{\fontsize{7pt}{7pt}\selectfont}
 
\begin{document}
   \title{Protoplanetary disks of \mbox{T\,Tauri} binary systems in the Orion Nebula Cluster\thanks{Based on observations collected at the European Organisation for Astronomical Research in the Southern Hemisphere, Chile. ESO Data ID: 074.C-0575.}}
   \titlerunning{Protoplanetary disks of T\,Tauri binaries in the ONC.}

   \author{S. Daemgen\inst{1}, S. Correia\inst{2}, \and M.~G. Petr-Gotzens\inst{1} }
   \authorrunning{Daemgen, Correia, \& Petr-Gotzens}

   \institute{European Southern Observatory,
     Karl-Schwarzschildstr. 2, 85748 Garching, Germany\\
              \email{sdaemgen@eso.org}
         \and
             Leibniz-Institut f\"ur Astrophysik Potsdam (AIP),
             An der Sternwarte 16, 14482 Potsdam, Germany
             }

   \date{}

  \abstract
   {}
   {We present a study of protoplanetary disks in spatially resolved low-mass binary stars in the well-known Orion Nebula Cluster (ONC) to assess the impact of binarity on the properties of circumstellar disks. This is currently the largest such study in a clustered high-stellar-density star-forming environment, as opposed to previous studies, which have mostly been focussed on the young, low-stellar-density Taurus association. We particularly aim to determine the presence of magnetospheric accretion and dust disks for each binary component, and measure the overall disk frequency.}
   {We carried out spatially resolved adaptive-optics-assisted observations to acquire near-infrared photometry and spectroscopy of 26 binaries in the ONC, and determine stellar parameters such as effective temperatures, spectral types, luminosities, and masses, as well as accretion properties and near-infrared excesses for the individual binary components. On the basis of our medium resolution $K$-band spectroscopy, we infer the presence of magnetospheric accretion around each binary component by measuring the strength of the Brackett-$\gamma$ emission. The accretion disk frequency among the ONC binaries is then estimated from Bayesian statistics. The observed disk signatures, measured accretion luminosities, and mass accretion rates are investigated with respect to the binary separation, mass ratios, and distance to the center of the ONC.}
   {A fraction of $40^{+10}_{-9}$\% of the binary components in the sample can be inferred to be \mbox{T\,Tauri} stars possessing an accretion disk. This is only marginally smaller than the disk fraction of single stars of $\sim$50\% in the ONC. We find that disks in wide binaries of $>$200\,AU separation are consistent with random pairing, while the evolution of circumprimary and circumsecondary disks is observed to be synchronized in close binaries (separations $<$200AU). Circumbinary disks appear to be unsuitable to explain this difference. Furthermore, we identify several mixed pairs of accreting and non-accreting components, suggesting that these systems are common and that there is no preference for either the more or less massive component to evolve faster. The derived accretion luminosities and mass accretion rates of the ONC binary components are of similar magnitude as those for both ONC single stars and binaries in the Taurus star-forming region. The paper concludes with a discussion of the (presumably weak) connection between the presence of inner accretion disks in young binary systems and the existence of planets in stellar multiples.}
   {}


   \maketitle
%

\section{Introduction}
One of the most important and explicit indicators of stellar youth are circumstellar disks, which seem to be a universal feature of the star-forming process. The existence of a stellar companion close to a disk-bearing star, however, has a complex impact on the evolution of a disk: it will be truncated to about a third of the binary separation, may become eccentric and warped, and the material in the disk may be heated and dynamically stirred \citep{art94,kle08,fra10,nel00}. Truncation, in particular, can explain observations of young low-mass binary components that have a systematically lower disk frequency than single stars with otherwise identical properties \citep{bou06,mon07,cie09}. Material that is usually transported inwards from the outer parts of the disk is missing, thus cannot replenish the inner disk and extend their disk lifetimes, in a way assumed for single star disks.

Primordial circumstellar disks are not only indicators of the formation of the star itself, but also contain the material for the formation of planetary systems. If disks in binaries evolve differently from those in single stars, then the properties of the population of planets found in binaries can reflect those differences. Today, more than 40 planets are known to orbit one of the components of a binary system, with binary separations of mostly $\ge$30\,AU \citep{mug09,egg10}. Interestingly, systems in which both components orbit their own planet are disproportionally rarely observed -- to the knowledge of the authors no system has been published so far. Although not free from observational selection effects, this might indicate that there are differences in the evolutions of the initial circumstellar material around the individual components of a binary system.

Another open issue is universality throughout the Galaxy.
Disk evolution is not determined entirely by the interaction with their host stars or stellar binary companions. The irradiation by nearby hot sources and the gravitational environment play an important role \citep{sca01}. While previous studies of the signatures of disks in binaries were mostly focused on nearby, loose, stellar associations such as Taurus and Ophiuchus \citep{whi01,har03,pra03,duc04,pat08,cie09}, there is evidence that most field stars are instead born in dense high-mass clusters such as the Orion Nebula Cluster \citep[ONC;][]{lad03}. 

As the nearest \citep[414$\pm$7\,pc;][]{men07} young \citep[$\sim$1\,Myr;][]{hil97} high-mass star forming cluster, the ONC has been targeted by many studies owing to its abundant circumstellar disk \citep[e.g.][]{hil98,lad00,da_10} and binary content \citep{pro94,pad97,pet98,sim99,khl06,rei07}, although a detailed survey of disks around binary components has yet to be performed. In this paper, we present the first spatially resolved near-infrared spectroscopic observations of a large sample of binaries in a clustered star-forming environment -- the Orion Nebula Cluster -- to investigate its circumstellar disk content and the binary component properties.

\section{Observations and data reduction}\label{sec:observations}
\subsection{Sample definition}
We observed 20 visual binaries in the ONC from the binary census of \citet{pet98a} and \citet{khl06}. Data for six additional binaries were taken from observations by \citet[\emph{in prep.}]{cor12}. All targets are listed in Table~\ref{tab:observations}. 
\begin{table*}
  \caption{Targets and observations}
  \label{tab:observations}
  \begin{center}
  \begin{tabular}{lcccccc}
    \hline\hline\\[-2ex]
    &
    dist. to &
    &
    \underline{I}maging or &
    Imaging &
    &
    Date \\
    Name\tablefootmark{a} &
    $\theta^1$\,Ori\,C [$^\prime$]&
    Obs. with&
    \underline{S}pectroscopy &
    Filters\tablefootmark{b} &
    SB\tablefootmark{c} &
    (UT) \\[0.5ex]
    \hline\\[-2ex]
    \object{[AD95]\,1468}             &  7.18   &   NACO   &    I\&S    &$JH$            &  no     & Feb 09, 2005                \\ 
    \object{[AD95]\,2380}             &  6.92   &   NACO   &    I\&S    &$JH$            &  no     & Jan 06, 2005                \\ 
    \object{JW\,235}                  &  6.85   &   NACO   &    I\&S    &$JH$            &  no     & Dec 19, 2004                \\ 
    \object{JW\,260}                  &  9.00   &   NACO   &    I\&S    &$JH$            &  SB1    & Dec 19, 2004                \\ 
    \object{JW\,519}                  &  0.80   &   NACO   &     I      &$JHK_\mathrm{s}$ &         &  Feb 07 \& 09, 2005          \\
    \object{JW\,553}                  &  0.52   &   NACO   &     S      &                &         & Dec 07, 2004                \\ 
    \object{JW\,566}                  &  7.17   &   NACO   &    I\&S    &$JH$            &         & Dec 08, 2004 \& Feb 17, 2005\\ 
    \object{JW\,598}                  &  0.71   &   NACO   &     S      &                &         & Dec 08, 2004                \\ 
    \object{JW\,648}                  &  1.00   &   NACO   &    I\&S    &$JHK_\mathrm{s}$ &         &  Feb 07 \& 17, 2005          \\
    \object{JW\,681}                  &  1.25   &   NACO   &     S      &                &  no     & Jan 06, 2005                \\ 
    \object{JW\,687}                  &  2.06   &   NACO   &    I\&S    &$JH$            &         & Feb 09, 2005 \& Dec 07, 2004\\ 
    \object{JW\,765}                  & 14.09   &   NACO   &     I      &$JH$            &  no     & Feb 07, 2005                \\ 
    \object{JW\,876}                  & 14.45   &   NACO   &     S      &                &  no     & Jan 08, 2005                \\ 
    \object{JW\,959}                  &  7.98   &   NACO   &     S      &                &  no     & Jan 08, 2005                \\ 
    \object{JW\,974}                  & 14.96   &   NACO   &     I      &$JH$            &  SB2?   & Nov 19, 2004                \\ 
    \object{[HC2000]\,73}             &  2.30   &   NACO   &     S      &                &         & Jan 05, 2005                \\ 
    \object{TCC\,15}                  &  0.50   &   NACO   &     S      &                &         & Dec 29, 2004                \\ 
    \object{TCC\,52}                  &  0.46   &   NACO   &     S      &                &         & Dec 08, 2004                \\ 
    \object{TCC\,55}                  &  0.49   &   NACO   &     S      &                &         & Jan 10 \& 11, 2005          \\ 
    \object{TCC\,97}                  &  0.42   &   NACO   &     I      &$JHK_\mathrm{s}$ &         &  Feb 07, 2005                \\
    \object{JW\,63}\tablefootmark{d}  &  8.95   &  GEMINI  &    I\&S    &$JHKL^\prime$    &         &  Feb 16 \& 19, 2008          \\
    \object{JW\,128}\tablefootmark{d} &  6.10   &  GEMINI  &    I\&S    &$JHKL^\prime$    &         &  Feb 17 \& 19, 2008          \\
    \object{JW\,176}\tablefootmark{d} &  4.67   &  GEMINI  &    I\&S    &$JHKL^\prime$    &  no     &  Feb 24 \& Mar 7, 2008       \\
    \object{JW\,391}\tablefootmark{d} &  2.81   &  GEMINI  &    I\&S    &$JHKL^\prime$    &         &  Feb 19, 2008                \\
    \object{JW\,709}\tablefootmark{d} &  3.53   &  GEMINI  &    I\&S    &$JHKL^\prime$    &         &  Feb 18 \& 20, 2008          \\
    \object{JW\,867}\tablefootmark{d} &  5.79   &  GEMINI  &    I\&S    &$JHKL^\prime$    &  no     &  Feb 23, 2008                \\
    \hline
  \end{tabular}
  \end{center}
\vspace{-3ex}
  \tablefoot{
      \tablefoottext{a} Identifiers are suitable for use with the simbad database ({\tt http://simbad.harvard.edu/simbad/}). References: \emph{AD95}: \citealt{ali95}; \emph{JW}: \citealt{jon88}; \emph{TCC}: \citealt{mcc94}; \emph{HC2000}: \citealt{hil00}.
      \tablefoottext{b} Photometry for this study was obtained in these filters and has been listed in Table~\ref{tab:systemparameters}. If fewer filters than the complete set of $JHK_\mathrm{s}$ has been indicated here, photometry has been added to Table~\ref{tab:systemparameters} from literature sources when available.
      \tablefoottext{c} Spectroscopic binary status.
      References: \citet{tob09,frs08}.
      \tablefoottext{d} From \citet[\emph{in prep.}]{cor12}.
  }
\end{table*}
 The projected separations range from 0.25 to 1.1\,arcsec, which corresponds to roughly 100--400\,AU at the distance of the ONC. Magnitude differences of the binary components range from 0.1 to $\sim$3\,mag in $H$ and $K_\mathrm{s}$-band.

All of our targets are members or very likely members of the ONC, which were mostly identified based on their proper motion \citep[and references therein]{hil97}. Targets without proper motion measurements ([HC2000]\,73, TCC\,15, TCC\,55) or those with low proper-motion membership probability (JW\,235, JW\,566, JW\,876) were confirmed to be young stars, thus likely members based on their X-ray activity \citep{hil00,rei07,get05}. No further information is available for two targets, [AD95]\,1468 and [AD95]\,2380, but our spectroscopy shows late spectral types at moderate luminosity and extinction for these and all other targets, ruling out foreground and background stars.
Common proper motion with the ONC and signs of youth combined with small angular separation of the components render it likely that all binaries are gravitationally bound. However, chance alignment of unrelated members of the Orion complex cannot be excluded.
A possible example might be TCC\,15, whose secondary shows almost no photospheric features but strong Br$\gamma$ and \ion{He}{I} emission, suggesting a highly veiled nature of this component. This makes it a candidate member of the Orion BN/KL region, which is in the line of sight of the ONC but slightly further away ($\sim$450\,pc) and probably younger \citep{men07}. However, the chance of finding an unrelated stellar component at the separation measured for this binary (1\farcs02) is only $W$=8\% given a projected density of $\sim$0.03\,stars/arcsec$^2$ in the center of the ONC \citep{pet98}. 
In the following, we treat TCC\,15 as a physical binary.

We also searched the literature for possible spectroscopic pairs among our binary components and found that among the 11 binaries that had been surveyed \citep{tob09,frs08}, two are spectroscopic binary candidates (see Table~\ref{tab:observations}). Both had already been excluded from our statistical analysis for other reasons (see discussion in Sect.~\ref{sec:brgammavsnirexcess} for JW\,260; JW\,974 was not observed with spectroscopy). Spectroscopic binarity cannot be excluded for the other targets, since no published spectroscopic binary surveys exist and our observations are of too low spectral resolution to detect spectroscopic pairs. In the following, we treat these binary components as single stars.

Table~\ref{tab:observations} lists all targets including the dates of their obervation, the instrument used, and the observational mode. Furthermore, we provide the projected distance to the massive and bright $\theta^1$\,Ori\,C system, which is at the center of the ONC. 

\subsection{NACO/VLT} Twenty of our 26 targets were observed with the NAOS-CONICA instrument \citep[NACO;][]{len03,rou03} at UT4/VLT in the time from November 2004 to February 2005. Of those, 16 were observed in spectroscopic mode and 11 in CONICA imaging mode (see Table~\ref{tab:observations}). All observations were executed using the adaptive optics system NAOS where the targets themselves could be used as natural guide stars. Depending on the brightness of a target, either an infrared or visual wavefront sensor was used.

\subsubsection{NACO Imaging}\label{sec:nacoimagingreduction} We employed the S13 camera of NACO with a pixel scale of 13.26\,mas/pixel and 13\farcs6$\times$13\farcs6 field of view. Each imaged target binary was observed in $J$ and $H$ filters and three of these additionally in $K_\mathrm{s}$. Observations of the same target in different filters were obtained consecutively to minimize the effects of variability. 
The FWHMs of the observations are typically 0\farcs{}075 in $J$, 0\farcs{}065 in $H$, and 0\farcs{}068 in $K_\mathrm{s}$-band. All observations were made in a two-offset dither pattern with a 5\arcsec\ pointing offset to allow for sky subtraction with total integration times varying between 60\,s and 360\,s per target and filter, depending on the brightness of the target components.

\begin{figure*}
  \centering
  \includegraphics[angle=0,width=0.98\textwidth]{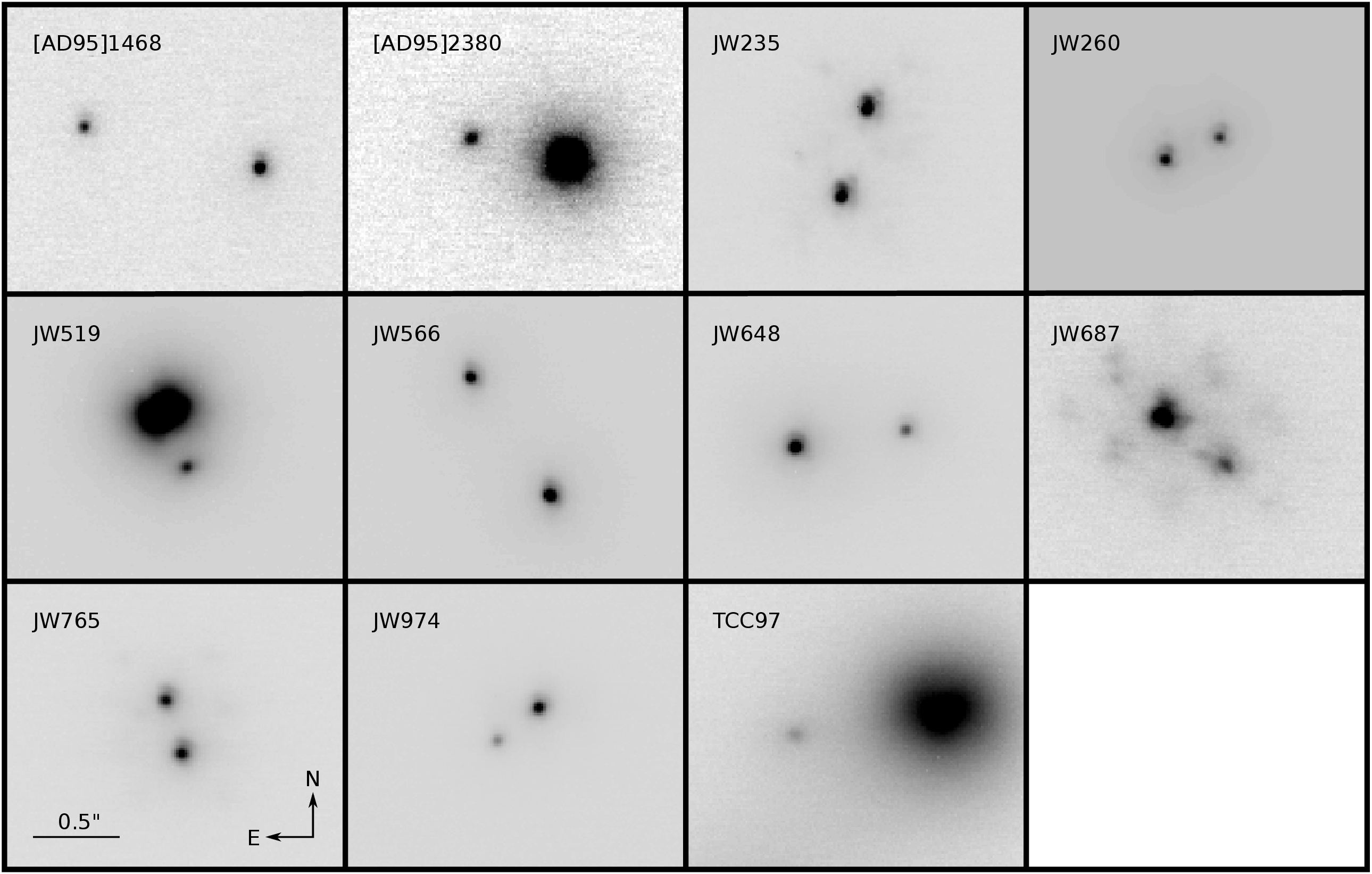}
  \caption{$H$-band images of all targets observed with NACO imaging. The intensity scale is linear and adapted to best depict both components.}
  \label{fig:imaging}
\end{figure*}

Imaging data were reduced with custom IRAF and IDL routines according to the following procedure. A low amplitude ($<10$\,ADU), roughly sinusodial horizontal noise pattern--probably 50\,Hz pick-up noise at read-out--was removed from all raw images by subtracting a row-median (omitting the central region containing the signal) from each column. We cropped each image to a 400$\times$400 pixel ($\equiv 5\farcs3\times 5\farcs3$) subregion of the original 1024$\times$1024 pix preserving all flux from each target multiple. For each target and filter there were at least two dithered images that undergo the same corrections, which were then used for sky subtraction. Bad pixel correction and flat fielding were applied, the latter using lamp flats taken with the NACO internal calibration unit. All reduced images per object and filter were then aligned and averaged. The fully reduced images in $H$-band are presented in Fig.~\ref{fig:imaging}.

\subsubsection{NACO Spectroscopy} NACO was used in grism-spectroscopy setup using the S27 camera and an 86\,mas wide slit. The wavelength coverage is $2.02$--$2.53$\,$\mu${}m with a spectral resolution of $R$$\sim$$1400$ and a resolution of 27\,mas/pix in the spatial direction. The slit was aligned with the binary separation vector to simultaneously obtain spectra of both stellar components. All binaries were observed in an ABBA nodding pattern with a 12\arcsec\ nod throw, performing several nod cycles for the faintest targets. Total exposure times per target were 120$-$2600\,s. Spectroscopic standards of B spectral type were observed with the same camera setup, close in time and at similar airmass to enable us to remove telluric features. These standard spectra were reduced and extracted in the same way as the target exposures.

Spectroscopic exposures were reduced with IDL and IRAF routines including flat fielding, sky subtraction, and bad pixel removal. Lamp flats taken with an internal flat screen were fitted along the dispersion axis and divided. The extraction and wavelength calibration of the reduced target and standard exposures were performed using the \emph{apextract} and \emph{dispcor} packages in \emph{IRAF}. The two-dimensional spectrum was traced with a fourth order polynomial and extracted through averaging over a 10 pixel-wide ($\sim$0\farcs27) aperture centered on the $\sim$4 pixel-wide (FWHM) trace of each target component. The subsequent wavelength calibration uses exposures of an argon arc lamp extracted in the same traces. All nodding exposures of the same target component were then aligned and averaged. The final extracted and calibrated spectra span a wavelength range of 20320--25440\,\AA\ with a resolution of 5\,\AA/pixel.

To allow for accurate telluric-line removal, intrinsic spectral features of the standard stars had to be removed. These were the \mbox{Brackett-$\gamma$} (Br$\gamma$, 21665\,\AA) and the \ion{He}{I} (21126\,\AA) absorption lines, where the latter was only observed for spectral types B0 and B1. The removal of in particular the Br$\gamma$ line is crucial because we aim to measure the equivalent widths of Br$\gamma$ emission in the target spectra. Therefore, the Br$\gamma$ line was carefully modeled according to the following scheme. Two telluric lines blend into the blue and red wings of Br$\gamma$ at the spectral resolution of our observations. We modeled these tellurics by averaging the fluxes in the Br$\gamma$ region (before telluric removal) of those target component spectra that showed neither absorption nor emission in Br$\gamma$. The telluric standard flux was divided by the thus generated local telluric model, allowing us to fit the remaining Br$\gamma$ absorption line with a Moffat-profile and remove its signature. The cleaned standard spectra were then divided by a blackbody curve of temperature $T_\mathrm{eff}$ according to their spectral type \citep{cox00} to obtain a pure telluric spectrum, convolved with the instrumental response of NACO.

We observed a wavelength-dependent mismatch of the wavelength calibrations output from \emph{IRAF/dispcor} between the target stars and the corresponding telluric standards of up to about one pixel. To guarantee a good match of the positions of the telluric features, we used the tellurics themselves to fine-tune the wavelength calibration of the standard spectra. A customized \emph{IDL} routine computes the local wavelength difference by means of cross correlation and corrects the mismatch accordingly. Since the tellurics are the strongest features in all our spectra, this method results in a good match between the telluric features of the target and reference, and telluric features could be removed reliably.

Flux uncertainties in the derived spectra as needed for the $\chi^2$ minimization method described in \S\ref{sec:fitting} were estimated from the fully reduced and extracted target spectra by performing local computation of standard deviations. The final reduced and extracted component spectra contain a noisy (signal-to-noise ratio (S/N) smaller than 20/pix) region redwards of $\sim$\,25000\,\AA\ caused by the low atmospheric transmission at these wavelengths. The region with $\lambda> 25120$\,\AA\ was excluded from any further evaluation. The set of reduced and extracted spectra consists of one spectrum for each target component in the spectral range of 20320\,\AA\ -- 25120\,\AA\ with a resolution of 5\,\AA/pixel. All final reduced spectra are displayed in Fig.~\ref{fig:spectroscopy}. The positions of the most prominent absorption and emission features are overplotted and listed in Table~\ref{tab:features}.

\begin{figure*}
  \begin{center}
    \includegraphics[angle=0,width=1.0\textwidth]{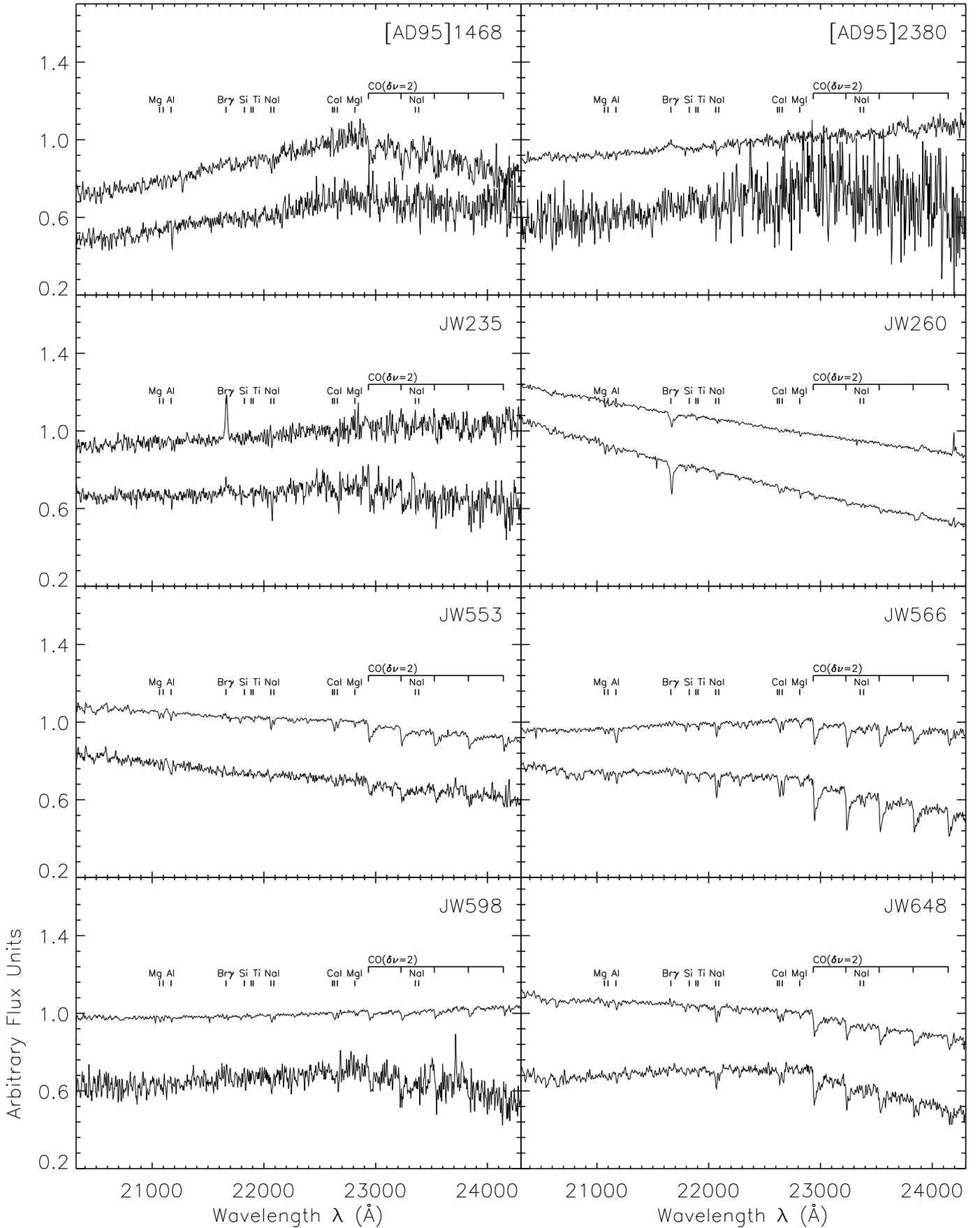}
    \caption{Spectra of the primary (top spectrum in each panel) and secondary component (bottom spectrum) of all targets observed with NACO spectroscopy (Gemini/NIFS spectra are displayed in \citet[\emph{in prep.}]{cor12}). Primary spectra are normalized at 2.2\,$\mu$m, and the secondaries are arbitrarily offset. The position of the most prominent lines in Table~\ref{tab:features} are indicated.}
    \label{fig:spectroscopy}
  \end{center}
\end{figure*}
\addtocounter{figure}{-1}
\begin{figure*}
  \begin{center}
    \includegraphics[angle=0,width=1.0\textwidth]{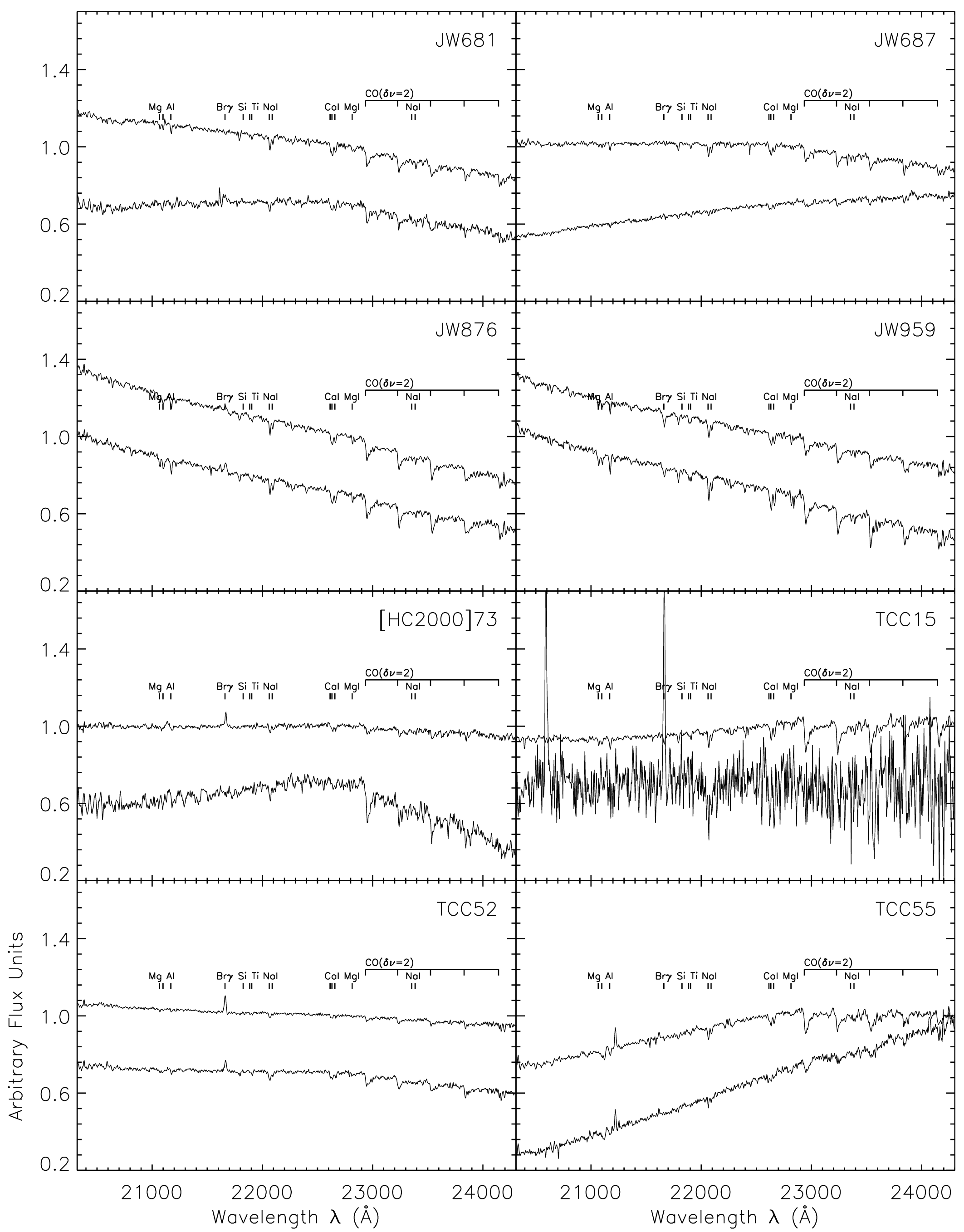}
    \caption{\it Continued}
  \end{center}
\end{figure*}

\begin{table}
  \caption{Spectral features identified in the observed spectra}
  \label{tab:features}
  \begin{center}
    \begin{tabular}{lccl}
      \hline\hline\\[-2ex]
      \multicolumn{1}{c}{$\lambda_\mathrm{c}$} & 
      \multicolumn{1}{c}{Width} &
      &
      \\
      \multicolumn{1}{c}{[\AA]} &
      \multicolumn{1}{c}{[\AA]} &
      \multicolumn{1}{c}{Species} &
      \multicolumn{1}{c}{Transition}\\[0.5ex]
      \hline\\[-2ex]
      20338.0     &         &   H$_2$         & $\nu = (1$--$0)\,S(2)$                              \\ 
      20587.0     &         &   \ion{He}{I}   & $2p^1P^o$--$2s^1S$                  \\
      21066.6     &         &   \ion{Mg}{I}   & $4f^3F^o_{2,3,4}$--$7g^3G^o_{3,4,5}$\\
      21098.8     &         &   \ion{Al}{I}   & $4p^2P^o_{1/2}$--$5s^2S_{1/2}$      \\
      21169.6     &         &   \ion{Al}{I}   & $4p^2P^o_{3/2}$--$5s^2S_{1/2}$      \\
      21218.0     &         &   H$_2$         & $\nu = (1$--$0)\,S(1)$              \\
      21661.2     &   56    &   \ion{H}{I}    & $n = 7-4\,\,(\mathrm{Br}\gamma)$    \\
      21785.7     &         &   \ion{Si}{I}   &                                     \\
      21825.7     &         &   \ion{Si}{I}   &                                     \\
      21903.4     &         &   \ion{Ti}{I}   & $a^5P_2$--$z^5D^o_3$                \\
      22062.4     &  116    &   \ion{Na}{I}   & $4p^2P^o_{3/2}$--$4s^2S_{1/2}$      \\
      22089.7     &    *    &   \ion{Na}{I}   & $4p^2P^o_{1/2}$--$4s^2S_{1/2}$      \\
      22614.1     &   91    &   \ion{Ca}{I}   & $4f^3F^o_2$--$4d^3D_1$              \\
      22631.1     &    *    &   \ion{Ca}{I}   & $4f^3F^o_3$--$4d^3D_2$              \\
      22657.3     &    *    &   \ion{Ca}{I}   & $4f^3F^o_4$--$4d^3D_3$              \\
      22814.1     &         &   \ion{Mg}{I}   & $4d^3D_{3,2,1}$--$6f^3F^o_{2,3,4}$  \\
      22935.3     &  170    &   \element[][12]{CO}     & $\nu=(2$--$0)$ band head            \\
      23226.9     &  170    &   \element[][12]{CO}     & $\nu=(3$--$1)$ band head            \\
      23354.8     &         &   \ion{Na}{I}   & $4p^2P^o_{1/2}$--$4d^2D_{3/2}$      \\
      23385.5     &         &   \ion{Na}{I}   & $4p^2P^o_{3/2}$--$4d^2D_{5/2}$      \\
      23524.6     &  170    &   \element[][12]{CO}     & $\nu=(4$--$2)$ band head            \\
      23829.5     &  170    &   \element[][12]{CO}     & $\nu=(5$--$3)$ band head            \\[0.5ex]
      \hline
    \end{tabular}
  \end{center}
\vspace{-3ex}
    \tablefoot{
      Identified features in our $K$-band spectra. Integration widths are listed here for all lines that have equivalent widths measured in this paper. Lines marked with asterisks blend into the line with the next shortest wavelength (the respective previous line in the list); these lines were integrated together. The transition information is composed from \citet{pra03} and \citet{kle86}.
      }
\end{table}

\subsection{NIFS-NIRI/Gemini North imaging and spectroscopy}
The observations with NIRI photometry and NIFS spectroscopy at Gemini North are described in \citet[\emph{in prep.}]{cor12}.
We used the reduced and extracted but non-telluric corrected spectra of Correia et al. (2011) and performed the telluric correction using the same method as for our NACO observations. Furthermore, since the NIFS spectra are of higher spectral resolution (R$\sim$5000) than both our NACO observations and the template spectra, we smoothed the NIFS spectra with a Gaussian kernel to a resolution of R$\sim$1400. These steps were included to guarantee that a coherent evaluation of all data is possible.
The six target spectra observed with NIFS that were included in this study will be presented in \citet[\emph{in prep.}]{cor12}.

\section{Results}\label{sec:results}

\subsection{Photometry and astrometry}\label{sec:photometry}
Relative aperture photometry for all targets observed with NACO was obtained by applying the PHOT task in the IRAF DAOPHOT package to each of the reduced binary star images. The aperture radius was varied from 2 to 20 pixels to find a possible convergence of the magnitude difference of primary and secondary. The differential photometry of most binaries converged for aperture sizes of 3 to 6 pixels allowing determination of their magnitude differences with uncertainties of $\Delta\mathrm{mag}\lesssim0.03$. For the binaries that did not converge but followed a monotonic decrease in their magnitude difference with aperture size, we assigned a value of $\Delta\mathrm{mag}$ by averaging the results for apertures of sizes between 3 and 6 pixels. The uncertainty was estimated according to the slope of each individual curve.

The DAOFIND task in DAOPHOT returned sufficiently accurate astrometry for all targets. The pixel data were transformed into physical angles and separations using the pixel scale of 0.013260\,arcsec/pixel and the rotation offset of $0^\circ$. The resulting relative photometry and astrometry are listed in Table~\ref{tab:systemparameters}.
\begin{table*}
  \caption{Relative photometry and astrometry of the observed binaries\tablefootmark{\dag}}
  \label{tab:systemparameters}
  \begin{center}
  \begin{tabular}{lr@{\,$\pm$\,}lr@{\,$\pm$\,}lr@{\,$\pm$\,}lr@{\,$\pm$\,}lcccc}
    \hline\hline\\[-2ex]
    &
    \multicolumn{2}{c}{$\Delta J$} &
    \multicolumn{2}{c}{$\Delta H$} &
    \multicolumn{2}{c}{$\Delta K_\mathrm{s}$} &
    \multicolumn{2}{c}{$\Delta K$} &
    sep\tablefootmark{a} &
    PA\tablefootmark{b} &
    &
    \\
    Name &
    \multicolumn{2}{c}{[mag]} &
    \multicolumn{2}{c}{[mag]} &
    \multicolumn{2}{c}{[mag]} &
    \multicolumn{2}{c}{[mag]} &
    [\arcsec] &
    [$^\circ$] &
    Ref. \\[0.5ex]
    \hline\\[-2ex]
    {[AD95]\,1468} &      0.75 & 0.10     &     0.66 & 0.02      &       0.10 & 0.05    & \multicolumn{2}{c}{} & 1.08            &  76.9         & T,1 \\
    {[AD95]\,2380} &\multicolumn{2}{c}{$\gtrsim$2.5\tablefootmark{c}}&     2.62 & 0.03      &       2.99 & 0.15    & \multicolumn{2}{c}{} & 0.59            &  77.6         & T,1 \\
    {JW\,235}      &      0.46 & 0.02     &     0.10 & 0.02      &       0.47 & 0.15    & \multicolumn{2}{c}{} & 0.35            & 163.6         & T,1 \\
    {JW\,260}      &      0.53 & 0.02     &     0.40 & 0.02      &       0.17 & 0.05    & \multicolumn{2}{c}{} & 0.35            & 292.2         & T,1 \\
    {JW\,519}\tablefootmark{e} &      2.8  & 0.2      &     2.8  & 0.2       &       2.6  & 0.1     & \multicolumn{2}{c}{} & 0.36            & 204.3         & T   \\
    {JW\,553}      &      2.1  & 0.3      &     3.2  & 0.3       &       3.19 & 0.10    & \multicolumn{2}{c}{} & 0.384$\pm$0.004 & 248.1$\pm$0.3 & 2   \\
    {JW\,566}      &      0.20 & 0.03     &     0.39 & 0.02      &       0.78 & 0.02    & \multicolumn{2}{c}{} & 0.86            &  33.8         & 5   \\
    {JW\,598}      & \multicolumn{2}{c}{} & \multicolumn{2}{c}{} & \multicolumn{2}{c}{} & \multicolumn{2}{c}{3.19} & 0.9         &               &  4  \\
    {JW\,648}      &      1.12 & 0.03     &     1.16 & 0.03      &       1.23 & 0.02    & \multicolumn{2}{c}{} & 0.68            & 278.7         & T   \\
    {JW\,681}      & \multicolumn{2}{c}{} & \multicolumn{2}{c}{} & \multicolumn{2}{c}{} & \multicolumn{2}{c}{1.53} & 1.09        & 214           & 3,4 \\
    {JW\,687}      &      1.63 & 0.05     &     1.15 & 0.07\tablefootmark{d}&       0.53 & 0.07    & \multicolumn{2}{c}{} & 0.47            & 232.6         & T   \\
    {JW\,765}      &      0.08 & 0.02     &     0.12 & 0.01      &       0.06 & 0.10    & \multicolumn{2}{c}{} & 0.33            &  16.5         & T,1 \\
    {JW\,876}      & \multicolumn{2}{c}{} & \multicolumn{2}{c}{} &       0.50 & 0.05    & \multicolumn{2}{c}{} & 0.49            &               & 1   \\
    {JW\,959}      & \multicolumn{2}{c}{} & \multicolumn{2}{c}{} &       0.07 & 0.03    & \multicolumn{2}{c}{} & 0.34            &               & 1   \\
    {JW\,974}      &      1.16 & 0.04     &     1.26 & 0.04      &       1.41 & 0.05    & \multicolumn{2}{c}{} & 0.32            & 128.7         & T,1 \\
    {[HC2000]\,73}  &      0.48 & 0.09     &     1.13 & 0.06      & \multicolumn{2}{c}{} &       1.52 & 0.07    & 0.71            & 266.2$\pm$0.5 & 4   \\
    {TCC\,15}      &      2.5  & 0.3      &     3.3  & 0.3       &       3.65 & 0.1     & \multicolumn{2}{c}{} & 1.022$\pm$0.004 & 288.2$\pm$0.3 & 2 \\
    {TCC\,52}      &      1.61 & 0.03     &     1.56 & 0.03      &       1.64 & 0.02    & \multicolumn{2}{c}{} & 0.52            &  39.24        & 5   \\
    {TCC\,55}       &      1.0  & 0.1\tablefootmark{f}&     2.2  & 0.3       &       1.26 & 0.10    & \multicolumn{2}{c}{} & 0.256$\pm$0.004 & 153.1$\pm$0.3 & 2   \\
    {TCC\,97}       &      1.8  & 0.15     &     1.8  & 0.15      &       1.4  & 0.2     & \multicolumn{2}{c}{} & 0.88            &  98.5         & T   \\
  \hline
  \end{tabular}
  \end{center}
\vspace{-3ex}
  \tablefoot{
      \tablefoottext{\dag} Reproduced here are only the targets observed with NACO. The photometry of the six additional targets observed with Gemini/NIRI can be found in \citet[\emph{in prep.}]{cor12}.
      \tablefoottext{a} Uncertainty in the separation: $\Delta\mathrm{sep}=0\farcs01$ unless otherwise noted.
      \tablefoottext{b} Position angle uncertainty: $\Delta\mathrm{PA}=0.5^\circ$ unless otherwise noted.
      \tablefoottext{c} The companion to {[AD95]\,2380}A is detected with less than 3$\sigma$ significance in $J$-band. The number quoted here is a lower limit.
      \tablefoottext{d} This number is an average of two independent measurements that differ by 0.16 magnitudes.
      \tablefoottext{e} The elongated shape of {JW\,519}'s primary suggests the primary to be binary itself. The photometry for {JW\,519}, however, is based on the assumption of a single central object since separate components cannot be identified.
      \tablefoottext{f} Reanalysis of MAD data published in \citet{bou08}.
      \tablebib{
        (T) This paper; (1) \citealt{khl06}; (2) \citealt{bou08}; (3) \citealt{rei07}; (4) \citealt{pet98a}; (5) ESO archival data \mbox{074.C-0637(A)}. Reduction and photometry are described in Sects.~\ref{sec:nacoimagingreduction} and \ref{sec:photometry}.
      }
  }
\end{table*}

The procedure was successful for all binaries observed with NACO, although we found peculiarities for two targets. The image of the brighter component of {JW\,519} has an elongated shape (see Fig.~\ref{fig:imaging}) that is not seen in the companion point spread function (PSF). We therefore conclude that we have found evidence of an unresolved third component in the {JW\,519} system. Unfortunately, the close separation of the system does not allow us to determine any of the separate parameters of the individual components. The photometry of {JW\,519} in Tables~\ref{tab:systemparameters} and \ref{tab:componentmagnitudes}, as well as all further evaluation, therefore treats this likely triple as a binary.
The primary of {TCC\,97} is surrounded by a proplyd that was first identified by \citet{ode93}. Since this feature is detected in our images, our aperture photometry averages only the smallest useful apertures of 3--5 pixel radius (instead of 3--6) to exclude as much of the disk flux as possible.

A special photometry routine was applied to the target binaries {JW\,553}, {[HC2000]\,73}, and {TCC\,55}, which used photometry of several reference stars in the same exposure. In particular for {JW\,553} and {TCC\,55}, this procedure leads to more accurate photometry in the environment of strong nebulosity and high stellar density close to the cluster center. The photometry of {JW\,553} and {TCC\,55} uses fully reduced $J$, $H$, and $K$-band mosaics of the Trapezium region \citep{bou08} observed with the MCAO demonstrator \emph{MAD} \citep{mar07}. A reference PSF was computed from 4--8 stars within $\sim$9\arcsec\ of the target binary. With this PSF, we obtained instrumental photometry for all stars in the vicinity of the target using the IRAF daophot package. Observed apparent magnitudes of the reference stars together with our measured relative magnitudes were then used to derive the apparent magnitudes of the target binary components. Similarly, PSF photometry of {[HC2000]\,73} was derived using ADONIS \citep{pet98a} exposures making use of the PSF and apparent magnitude of the nearby star $\theta^2$\,Ori\,A \citep{mue02}.

Most photometry in this paper was obtained in the NACO and 2MASS $JHK_\mathrm{s}$ filter systems. Since these two filter sets are very similar, no transformations had to be applied. Some photometry, however, was performed in different filters. We checked the compatibility of the 2MASS $K_\mathrm{s}$ and the ADONIS and MAD $K$ filters used for some of the systems and relative magnitudes (see Tables~\ref{tab:systemparameters} and \ref{tab:integratedphotometry}).
 Comparing 39 young stars in the ONC (with similar properties as the target sample) for which both ADONIS $K$ and $K_\mathrm{s}$ photometry are available \citep{pet98a}, we found an average offset of only 0.01$\pm$0.23\,mag between $K$ and $K_\mathrm{s}$. The Gemini/NIRI $K$-band photometry of late-type stars is compatible with that of the $K_\mathrm{s}$-band with $K-K_\mathrm{s}=0\pm0.05$\,mag \citep{dae07}. We therefore did not apply any corrections to our $K$-band magnitudes. Some of the relations used in Sect.~\ref{sec:AVCTTS} (i.e.\ the dwarf, giant, and CTT locuses) were transformed to the 2MASS system using the relations in \citet{car01}. In the following, we thus assume that all photometry is compatible with the 2MASS filter system. 

The relative photometric and astrometric results are given in Table~\ref{tab:systemparameters}. Together with literature values for the integrated photometry in Table~\ref{tab:integratedphotometry}, component magnitudes were derived, which are listed in Table~\ref{tab:componentmagnitudes}.

\begin{table*}
  \caption{Non-resolved photometry of the observed binaries}
  \label{tab:integratedphotometry}
  \begin{center}
  \begin{tabular}{lr@{\,$\pm$\,}lr@{\,$\pm$\,}lr@{\,$\pm$\,}lr@{\,$\pm$\,}lc}
    \hline\hline\\[-2ex]
    &
    \multicolumn{2}{c}{$J^\mathrm{sys}$} &
    \multicolumn{2}{c}{$H^\mathrm{sys}$} &
    \multicolumn{2}{c}{$K_\mathrm{s}^\mathrm{sys}$} &
    \multicolumn{2}{c}{$K^\mathrm{sys}$} &
    \\
    Name &
    \multicolumn{2}{c}{[mag]} &
    \multicolumn{2}{c}{[mag]} &
    \multicolumn{2}{c}{[mag]} &
    \multicolumn{2}{c}{[mag]} &
    Ref. \\[0.5ex]
    \hline\\[-2ex]
    {[AD95]\,1468} &     13.92 & 0.11     &     12.48 & 0.18     &     11.23 & 0.08     & \multicolumn{2}{c}{} & 3\tablefootmark{a}\\
    {[AD95]\,2380} &     13.9  & 0.02     &     11.39 & 0.03     &      9.88 & 0.02     & \multicolumn{2}{c}{} & 3\tablefootmark{a}\\
    {JW\,235}      &     12.10 & 0.15     &     11.11 & 0.15     &     10.50 & 0.15     & \multicolumn{2}{c}{} & 2      \\
    {JW\,260}      &      8.19 & 0.15     &      7.60 & 0.15     &      7.23 & 0.15     & \multicolumn{2}{c}{} & 2      \\
    {JW\,519}      &     12.07 & 0.01     &     11.07 & 0.01     &     10.61 & 0.01     & \multicolumn{2}{c}{} & 1      \\
    {JW\,553}      &     10.56 & 0.10     &      9.49 & 0.05     & \multicolumn{2}{c}{} &     9.05 & 0.10      & \tablefootmark{b} \\
    {JW\,566}      &     11.49 & 0.04     &      9.97 & 0.05     &      8.86 & 0.03     & \multicolumn{2}{c}{} & 3\tablefootmark{a}\\
    {JW\,598}      &     10.87 & 0.04     &      9.57 & 0.05     &      8.93 & 0.11     & \multicolumn{2}{c}{} & 2      \\
    {JW\,648}      &     10.97 & 0.03     &      9.91 & 0.02     &      9.28 & 0.01     & \multicolumn{2}{c}{} & 1      \\
    {JW\,681}      &     12.78 & 0.01     &     11.82 & 0.01     &     11.07 & 0.11     & \multicolumn{2}{c}{} & 1      \\
    {JW\,687}      &     11.94 & 0.01     &     10.50 & 0.01     &      9.70 & 0.03     & \multicolumn{2}{c}{} & 1      \\
    {JW\,765}      &     11.76 & 0.15     &     11.04 & 0.15     &     10.81 & 0.15     & \multicolumn{2}{c}{} & 2      \\
    {JW\,876}      &      9.31 & 0.02     &      8.45 & 0.03     &      8.03 & 0.02     & \multicolumn{2}{c}{} & 3\tablefootmark{a}\\
    {JW\,959}      &      9.36 & 0.02     &      8.84 & 0.03     &      8.63 & 0.02     & \multicolumn{2}{c}{} & 3\tablefootmark{a}\\
    {JW\,974}      &     12.42 & 0.02     &     11.78 & 0.03     &     11.41 & 0.02     & \multicolumn{2}{c}{} & 3\tablefootmark{a}\\
    {[HC2000]\,73} &     12.59 & 0.05     &     11.72 & 0.03     &     10.99 & 0.03     & \multicolumn{2}{c}{} & \tablefootmark{b} \\
    {TCC\,15}      &     12.96 & 0.03     &     11.14 & 0.01     &     10.25 & 0.01     & \multicolumn{2}{c}{} & 1      \\
    {TCC\,52}      &      8.64 & 0.04     &      7.56 & 0.04     &      6.72 & 0.04     & \multicolumn{2}{c}{} & 2      \\
    {TCC\,55}      &     15.10 & 0.14     &     13.27 & 0.14     & \multicolumn{2}{c}{} &    11.15 & 0.14      & \tablefootmark{b} \\
    {TCC\,97}      &     13.13 & 0.03     &     12.44 & 0.02     &     11.77 & 0.01     & \multicolumn{2}{c}{} & 1      \\
  \hline
  \end{tabular}
  \end{center}
\vspace{-3ex}
  \tablefoot{
      \tablefoottext{a} If no other reference could be found \emph{and} if the distance to $\theta^1$\,Ori\,C is larger than $5\arcmin$, 2MASS values were used.
      \tablefoottext{b} System magnitudes were derived from the component magnitudes (Table~\ref{tab:componentmagnitudes}).
      \tablebib{
        (1) \citealt{mue02}; (2) \citealt{car01a}; (3) 2MASS, \citealt{cut03}.
      }
} 
\end{table*}

\begin{table}
  \caption{Individual component apparent magnitudes\tablefootmark{\dag}}
  \label{tab:componentmagnitudes}
  \begin{center}
    \begin{tabular}{lcr@{$\,\pm\,$}lr@{$\,\pm\,$}lr@{$\,\pm\,$}lr@{$\,\pm\,$}l}
      \hline\hline\\[-2ex]
      & 
      &
      \multicolumn{2}{c}{$J$} &
      \multicolumn{2}{c}{$H$} &
      \multicolumn{2}{c}{$K_\mathrm{s}$} \\
      \multicolumn{1}{c}{Name} &
       &
      \multicolumn{2}{c}{[mag]} &
      \multicolumn{2}{c}{[mag]} &
      \multicolumn{2}{c}{[mag]} \\[0.5ex]
      \hline\\[-2ex]
{}[AD95]\,1468  & A &         14.36 & 0.11         &         12.95 & 0.18         &         11.93 & 0.08                 \\
                & B &         15.11 & 0.13         &         13.61 & 0.18         &         12.03 & 0.08                 \\[0.5ex]
{}[AD95]\,2380  & A & \multicolumn{2}{c}{$\cdots$} &         11.48 & 0.03         &          9.95 & 0.02                 \\
                & B & \multicolumn{2}{c}{$\cdots$} &         14.10 & 0.04         &         12.94 & 0.14                 \\[0.5ex]
JW\,235         & A &         12.65 & 0.15         &         11.81 & 0.15         &         11.04 & 0.16                 \\ 
                & B &         13.11 & 0.15         &         11.91 & 0.15         &         11.51 & 0.18                 \\[0.5ex]
JW\,260         & A &          8.71 & 0.15         &          8.17 & 0.15         &          7.90 & 0.15                 \\ 
                & B &          9.24 & 0.15         &          8.57 & 0.15         &          8.07 & 0.15                 \\[0.5ex]
JW\,519         & A &         12.15 & 0.02         &         11.15 & 0.02         &         10.70 & 0.01                 \\
                & B &         14.95 & 0.19         &         13.95 & 0.19         &         13.30 & 0.09                 \\[0.5ex]
JW\,553         & A &         10.66 & 0.12         &          9.55 & 0.06         &          9.11 & 0.04\tablefootmark{a}\\
                & B &         13.18 & 0.09         &         12.70 & 0.06         &         12.21 & 0.13\tablefootmark{a}\\[0.5ex]
JW\,566         & A &         12.15 & 0.04         &         10.54 & 0.05         &          9.29 & 0.03                 \\
                & B &         12.35 & 0.04         &         10.93 & 0.05         &         10.07 & 0.03                 \\[0.5ex]
JW\,648         & A &         11.30 & 0.03         &         10.23 & 0.02         &          9.58 & 0.01                 \\
                & B &         12.42 & 0.04         &         11.39 & 0.03         &         10.81 & 0.02                 \\[0.5ex]
JW\,687         & A &         12.16 & 0.01         &         10.82 & 0.02         &         10.22 & 0.04                 \\ 
                & B &         13.79 & 0.04         &         11.97 & 0.05         &         10.75 & 0.05                 \\[0.5ex]
JW\,765         & A &         12.47 & 0.15         &         11.73 & 0.15         &         11.53 & 0.16                 \\ 
                & B &         12.55 & 0.15         &         11.85 & 0.15         &         11.59 & 0.16                 \\[0.5ex]
JW\,876         & A & \multicolumn{2}{c}{$\cdots$} & \multicolumn{2}{c}{$\cdots$} &          8.56 & 0.03                 \\ 
                & B & \multicolumn{2}{c}{$\cdots$} & \multicolumn{2}{c}{$\cdots$} &          9.06 & 0.04                 \\[0.5ex]
JW\,959         & A & \multicolumn{2}{c}{$\cdots$} & \multicolumn{2}{c}{$\cdots$} &          9.35 & 0.02                 \\ 
                & B & \multicolumn{2}{c}{$\cdots$} & \multicolumn{2}{c}{$\cdots$} &          9.42 & 0.03                 \\[0.5ex]
JW\,974         & A &         12.74 & 0.02         &         12.08 & 0.03         &         11.67 & 0.02                 \\ 
                & B &         13.90 & 0.04         &         13.34 & 0.04         &         13.08 & 0.04                 \\[0.5ex]
{}[HC2000]\,73  & A &         13.13 & 0.06         &         12.05 & 0.03         &         11.23 & 0.03                 \\ 
                & B &         13.61 & 0.07         &         13.18 & 0.05         &         12.75 & 0.06                 \\[0.5ex]
TCC\,15         & A &         13.06 & 0.04         &         11.19 & 0.02         &         10.29 & 0.01                 \\
                & B &         15.56 & 0.27         &         14.49 & 0.29         &         13.94 & 0.10                 \\[0.5ex]
TCC\,52         & A &          8.86 & 0.04         &          7.79 & 0.04         &          6.94 & 0.04                 \\
                & B &         10.47 & 0.05         &          9.35 & 0.05         &          8.58 & 0.04                 \\[0.5ex]
TCC\,55         & A &         14.45 & 0.08         &         12.67 & 0.17         &         10.96 & 0.03\tablefootmark{a}\\
                & B &         15.45 & 0.10         &         14.84 & 0.31         &         12.22 & 0.08\tablefootmark{a}\\[0.5ex]
TCC\,97         & A &         13.32 & 0.04         &         12.63 & 0.03         &         12.03 & 0.04                 \\
                & B &         15.12 & 0.13         &         14.43 & 0.13         &         13.43 & 0.16                 \\[0.5ex]
    \hline     
    \end{tabular}
  \end{center}
\vspace{-3ex}
    \tablefoot{
      \tablefoottext{\dag} The photometry of the six targets observed with Gemini/NIRI is listed in \citet[\emph{in prep.}]{cor12}.
      \tablefoottext{a} These values are derived from $K$-band system magnitudes instead of $K_\mathrm{s}$ (see Table~\ref{tab:integratedphotometry}).
      }
\end{table}

\subsection{Color-color diagram, extinctions, and color-excess}\label{sec:AVCTTS}
\begin{figure}
  \centering
  \includegraphics[angle=0,width=0.48\textwidth]{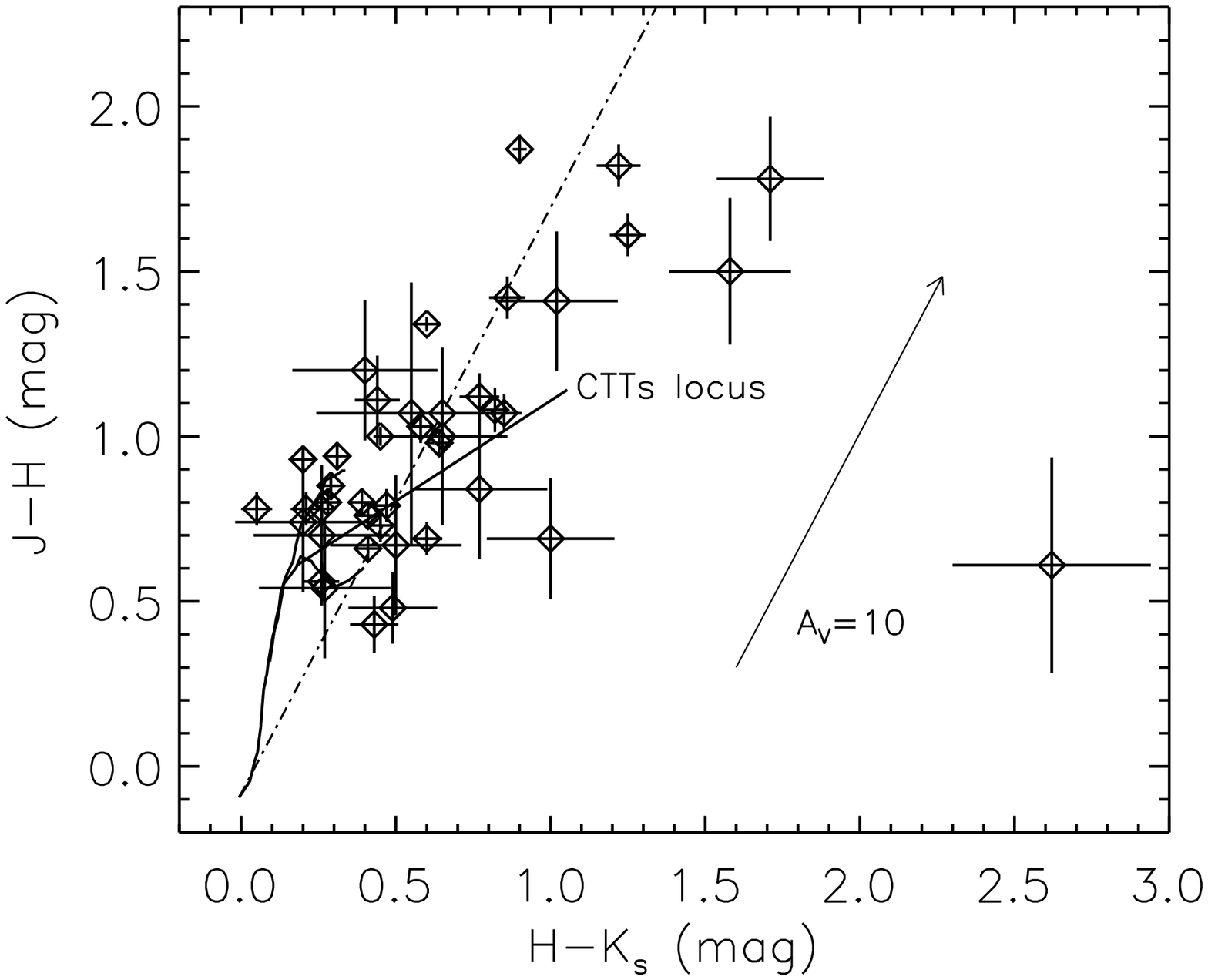}
  \caption{Color-color diagram of the target binary components. The CTTs locus \citep{mey97}, the dwarf and giant locus \citep{bes88}, and a reddening vector of 10\,mag length are overplotted, after conversion to the 2MASS photometric system \citep{car01}. Targets to the right of the dash-dotted line have an IR color excess.}
  \label{fig:colorcolor}
\end{figure}

Using the component magnitudes of Table~\ref{tab:componentmagnitudes}, we composed a ($H$$-$$K_\mathrm{s}$)-($J$$-$$H$) color-color diagram (Fig.~\ref{fig:colorcolor}).
 We compared our data with the loci of dwarfs and giants \citep{bes88} and the location of classical \mbox{T\,Tauri} stars \citep[CTTs locus;][]{mey97} that had both been converted to the 2MASS photometric system. Extinctions were derived by dereddening to the CTTs locus along the interstellar reddening vector \citep{coh81} and are listed in Table~\ref{tab:componentparameters}. 

Most targets are located in the region accessible from the dwarf and CTTs loci by adding extinction. However, three groups of targets have rather peculiar locations in the diagram: \emph{i)}~two of the targets with among the smallest of $H$$-$$K_\mathrm{s}$ values, the secondaries of {JW\,63} and {JW\,176}, have no intersection with either the dwarf or giant locus along the dereddening direction. \emph{ii)}~There are some targets (the secondaries to {JW\,553}, {[HC2000]\,73}, and {TCC\,97}) that are significantly below the CTTs locus. Targets if this location in the color-color diagram have been observed before \citep[see e.g.\ Fig.~20 in][]{rob10}. \emph{iii)}~The outlier {TCC\,55}B in the bottom-right of the plot.
Most of these peculiar locations can be explained by the intrinsic photometric variability of young stars of $\sim$0.2\,mag \citep{car01a} and the fact that photometry was not taken simultaneously. Furthermore, colors are known to depend on the inclination of a possible disk \citep{rob06} -- a parameter that cannot be determined with our data. All peculiar objects were assigned an extinction of 0. Using expected dwarf colors from \citet{bes88}, we then derived the color excesses $E_{J-H}=(J-H)_\mathrm{obs}-(J-H)_0$ and $E_{H-K_\mathrm{s}}$ for all objects (Table~\ref{tab:componentparameters}). 

We compared our extinction measurements with optical data from \citet{pro94}. Their sample contains six spatially resolved binaries that are also part of our sample (JW\,553, JW\,598, JW\,648, JW\,681, JW\,687, TCC\,15\footnote{\citet{pro94} list the brighter component in $V$ as the primary of binary. We show, however, that it is of later spectral type than its companion. Accordingly, our designation swaps both components with respect to the \citeauthor{pro94} paper.}). 
Using the observed $V\!-\!I$ colors and an estimate of $(V\!-\!I)_0$ derived from our measured effective temperatures
and the 1Myr stellar evolutionary model  of  \citet{bar98}, we calculated extinctions as $A_V \approx [(V-I) - (V-I)_0]/0.4$ \citep{bes88}.
The results are consistent with our near-infrared extinction measurements usually to within $\sim1.0$\,mag, except for JW\,687\,A, where we find a substantially larger $A_V$ from the optical data than from our NIR measurements. This discrepancy for this target, however, could be explained by its large veiling value (optical veiling can reduce the measured $V\!-\!I$ color and thus the derived extinction).

\subsection{Spectral types and veiling}\label{sec:fitting}
\begin{figure}
  \centering
  \includegraphics[width=0.48\textwidth]{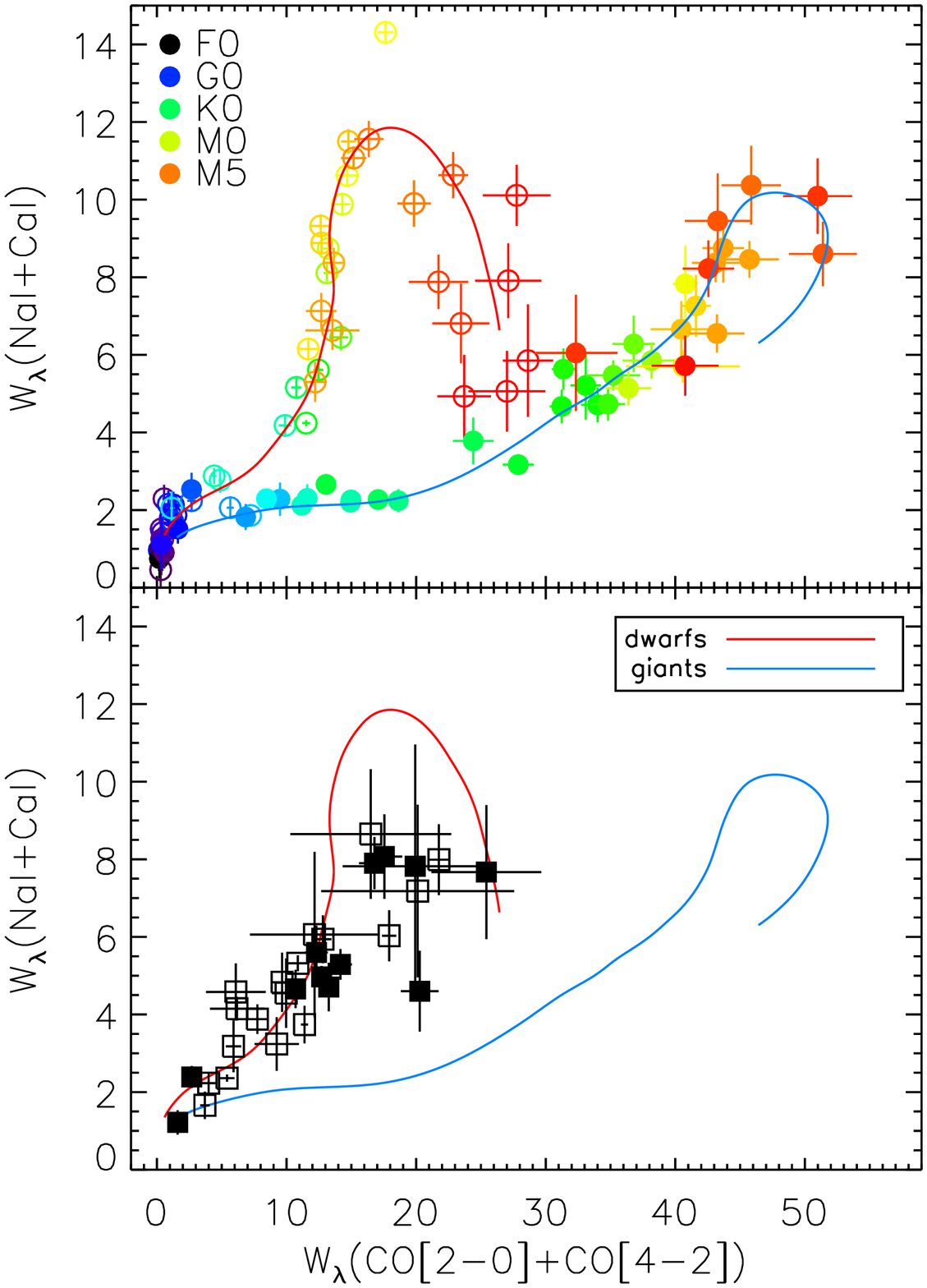}
  \caption{\label{fig:ew1}\emph{Top:} The equivalent widths of \ion{Na}{I}+\ion{Ca}{I} versus CO[2-0]+CO[4-2] as an indicator of T$_\mathrm{eff}$ and $\log g$ for dwarf (open circles) and giant (filled circles) template spectra from the IRTF spectral library. 
Spectral types are color-coded according to the legend in the upper left. 
The red and blue lines guide the eye to the two relations. \emph{Bottom:} The same plot including the derived dwarf and giant relations (red and blue curve) as in the top panel, overplot with the ONC target components. Open symbols show targetsted with high veiling $r_K>0.2$, filled symbols $r_K<0.2$.
Two targets, {TCC\,15}B and {[AD95]\,2380}B, are off bounds at high $W_\lambda$(\ion{Na}{I}+\ion{Ca}{I}).
}
\end{figure}

Owing to the young age of the targets in our sample, the stars are still contracting, hence their surface gravities ($\log g$) are lower than those of main sequence stars. However, to find appropriate templates for spectral classification and the estimation of both the visual extinction $A_V$ and near-infrared continuum excess in $K$-band (veiling, $r_K$), one needs to find templates with physical conditions as close as possible to the pre-main sequence stars in this sample. Since no comprehensive catalog of pre-main sequence spectra at a spectral resolution of $R=1400$ or higher exists, we compared our target spectra to those of dwarfs and giants using a method similar to that shown in \citet{pra03}. We measured the equivalent widths of the $T_\mathrm{eff}$ and $\log g$-sensitive photospheric features \ion{Na}{I}, \ion{Ca}{I}, \element[][12]{CO}(2-0), and \element[][12]{CO}(4-2) (wavelengths see Table~\ref{tab:features}) for all target component spectra as well as dwarf and giant spectra from the IRTF spectral libary \citep{ray09,cus05} in a spectral range between F2 and M9. Fig.~\ref{fig:ew1} demonstrates that dwarf spectra are more suitable templates for our sample than giant stars in the same spectral range.
We note that veiling reduces the equivalent widths of spectral lines according to $W_\lambda^\mathrm{measured} = W_\lambda/(1+r_K)$ (see also Sect.~\ref{sec:WBr}) and points in Fig.~\ref{fig:ew1} would move towards the origin if the veiling is high. It would therefore be necessary to only compare targets with small $r_K$, which is only possible after the veiling was determined using the templates. However, we see that not only the distribution of $r_K<0.2$ targets but all targets are well-congruent with the dwarf locus, indicating that dwarfs are suitable for the determination of the stellar parameters of our pre-main sequence stars. We also considered intermediate solutions between dwarfs and giants as possible templates. Despite the possibly closer match in $\log g$ of pre-main sequence stars, no improvement in the resulting match with our target equivalent widths could be seen. Thus, in order to minimize additional noise sources, we used dwarf templates for the evaluation that we now describe.

Spectral types, spectroscopic extinctions, and veiling were simultaneously determined by a $\chi^2$ minimization method that modifies the template spectra according to
\begin{equation}\label{eq:F}
  F^*(\lambda) = \left(\frac{F_\mathrm{phsph}(\lambda)}{c} + k\right)e^{-\tau_\lambda}
\end{equation}
\citep[cf.][]{pra03} with $\tau_\lambda=(0.522/\lambda)^{1.6}A_V^\mathrm{spec}$, where $F_\mathrm{phsph}$ is the photospheric flux of the templates and $k$ is the $K$-band excess in units of $F_\mathrm{phsph}(2.2\mu\mathrm{m})/c$, i.e. the excess over the photospheric flux at 2.2\,$\mu$m normalized with a constant $c$. We introduce the extinction variable $A_V^\mathrm{spec}$, since it is typically not identical to the photometric extinction $A_V$. The excess $k$ is assumed to not vary strongly with wavelength within the limits of $K$-band, i.e.\ it is 
$k(\lambda)=\mathrm{const}$. Although some of the targets show evidence of a slightly stronger excess towards the red edge of the $K$-band, the theoretically slightly poorer fit of line depth at the long wavelength end does not have a strong impact on the resulting $k$ since it is determined from the best fit to the entire wavelength range of our spectra. For each template spectral type, the three variables $A_V^\mathrm{spec}$, $k$, and $c$ were modified with 120$-$160 steps each within a reasonable range of values and we found a minimum value of
\begin{equation}
  \chi^2 = \frac{1}{n-\mathrm{\it dof}-1}\sum_{i=1}^n \frac{(F_i - F^*_i)^2}{\Delta F_i^2} 
\end{equation}
where $n$ is the number of pixels in the spectrum, $\mathrm{\it dof}=3$ the degrees of freedom, and $F_i$, $\Delta F_i$, and $F^*_i$ the flux in the $i$-th pixel of the target spectrum, its measurement error, and the modified template spectrum, respectively. The minimum of the $\chi^2$ distribution was compared for different spectral types and the nine best-fitting solutions (spectral type along with the corresponding optimal combination of $A_V^\mathrm{spec}$, $k$, $c$) when examined by eye. Spectral types were selected by comparing the overall shape of the continuum and the strength of several photospheric absorption lines of the modified (eq.~\ref{eq:F}) best-fit templates. Uncertainties in the spectral type were estimated from the range of spectra that could possibly fit the data. This resulted in a typical uncertainty of one or two subclasses. The uncertainty in $k$ was determined from the error ellipse in a three-parameter $\chi^2$-minimization at $\chi^2_\mathrm{min}+3.5$ \citep{wal96}.

\begin{figure}
  \centering
  \includegraphics[angle=0,width=0.48\textwidth]{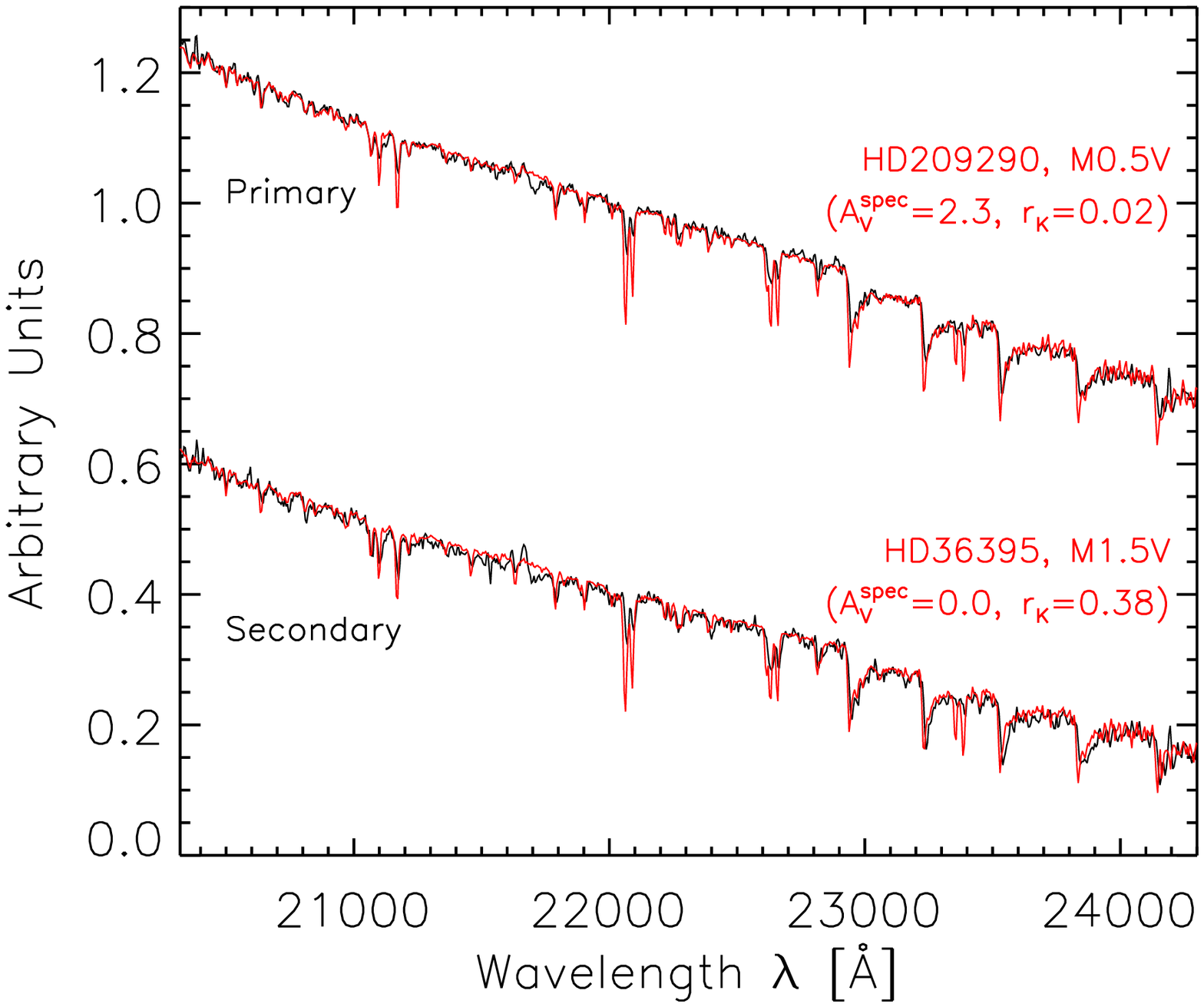}
  \caption{The result of the spectral template fitting for the primary and secondary component of JW\,876. The black curves show the spectrum of the primary and secondary component respectively, offset for clearer visibility. The red curves are the corresponding best-fit models from the IRTF spectral library, modified according to eq.~(\ref{eq:F}) with extinction and veiling values from Table~\ref{tab:componentparameters}.}
  \label{fig:bestfit}
\end{figure}

An example of the best-fit result for one of our target binaries is shown in Fig.~\ref{fig:bestfit}.
From the excess flux $k$, the $K$-band continuum excess $r_K$ is calculated to be
\begin{equation}
  r_K = \frac{k}{F_\mathrm{phsph}(2.2\mu\mathrm{m})/c}\quad.
\end{equation}
The results from the spectral fitting are summarized in Table~\ref{tab:componentparameters}.
\addtocounter{table}{1}
Some spectra do not show any photospheric features and no estimation of the spectral type was possible. These target components are marked with ellipses in the spectral type-dependent columns of Table~\ref{tab:componentparameters}.

Despite the wide range of extinction values derived from the photometric and spectroscopic determination, we did not force the $\chi^2$ minimization to match the photometric $A_V$ values from Sect.~\ref{sec:AVCTTS} for the following reason. In the $\chi^2$ fitting routine, the determination of the other parameters (spectral type, veiling, normalizing constant $c$) does not depend on the value chosen for $A_V$ as long as it has reasonable values. While spectral types are mainly determined by line ratios, veiling is sensitive to line depths; both signatures are not strongly influenced by extinction. However, forcing $A_V$ to a particular value typically produces poor fits and hence $A_V$ is kept as a free parameter.

We only use the photometric extinctions listed in Table~\ref{tab:componentparameters} for further evaluation. Exceptions are {JW\,681}, {JW\,876}, and {JW\,959} where no photometric extinctions could be measured and our best estimates for $A_V$ come from spectroscopy with an uncertainty determined from the fitting error ellipse that is similar to the uncertainty in $k$.

\subsection{Luminosity, effective temperature, and radius}
Luminosities $L_*$ of our target components were derived from bolometric magnitudes by applying bolometric corrections $BC\!_J$ \citep{har94} to the measured $J$-band magnitudes with a distance to Orion of $414\pm7$\,pc \citep{men07} and $J$-band extinctions of $A_J=0.27 A_V$ \citep{coh81}. The $J$-band was chosen to help us minimize the impact of hot circumstellar material, which mainly contributes flux at longer wavelengths ($K$ and $L$-band) i.e.\ closer to the maximum of the $T\sim1500$\,K blackbody emission from the inner dust rim \citep{mey97}. The resulting luminosities are listed in Table~\ref{tab:componentparameters}, along with effective temperatures $T_\mathrm{eff}$ derived from their spectral types and SpT--$T_\mathrm{eff}$ relations (earlier than M0: \citealt{sch82}; later or equal to M0: \citealt{luh03}). Luminosity uncertainties were propagated from the magnitude, extinction, and distance uncertainties but do not include any of the intrinsic variability of the targets since we cannot estimate the magnitude of the effect for any individual target component\footnote{The variability of a few of the target binaries was observed by \citet{car01a}, although they did not resolve the individual components.}. We estimate, however, an average impact of variability by varying $M_J$ by 0.2\,mag \citep{car01a} and rederiving the luminosity of each target. We observe a difference in the derived luminosities of up to 20\% with a sample median of 0.08\,dex. This is twice as large as the typical propagated random uncertainties. 
Stellar radii $R$ were then calculated from $L_*$ and $T_\mathrm{eff}$.

\subsubsection{The HR diagram: Ages and masses}\label{sec:HRdiagram}
\begin{figure}
  \centering
  \includegraphics[angle=0,width=0.48\textwidth]{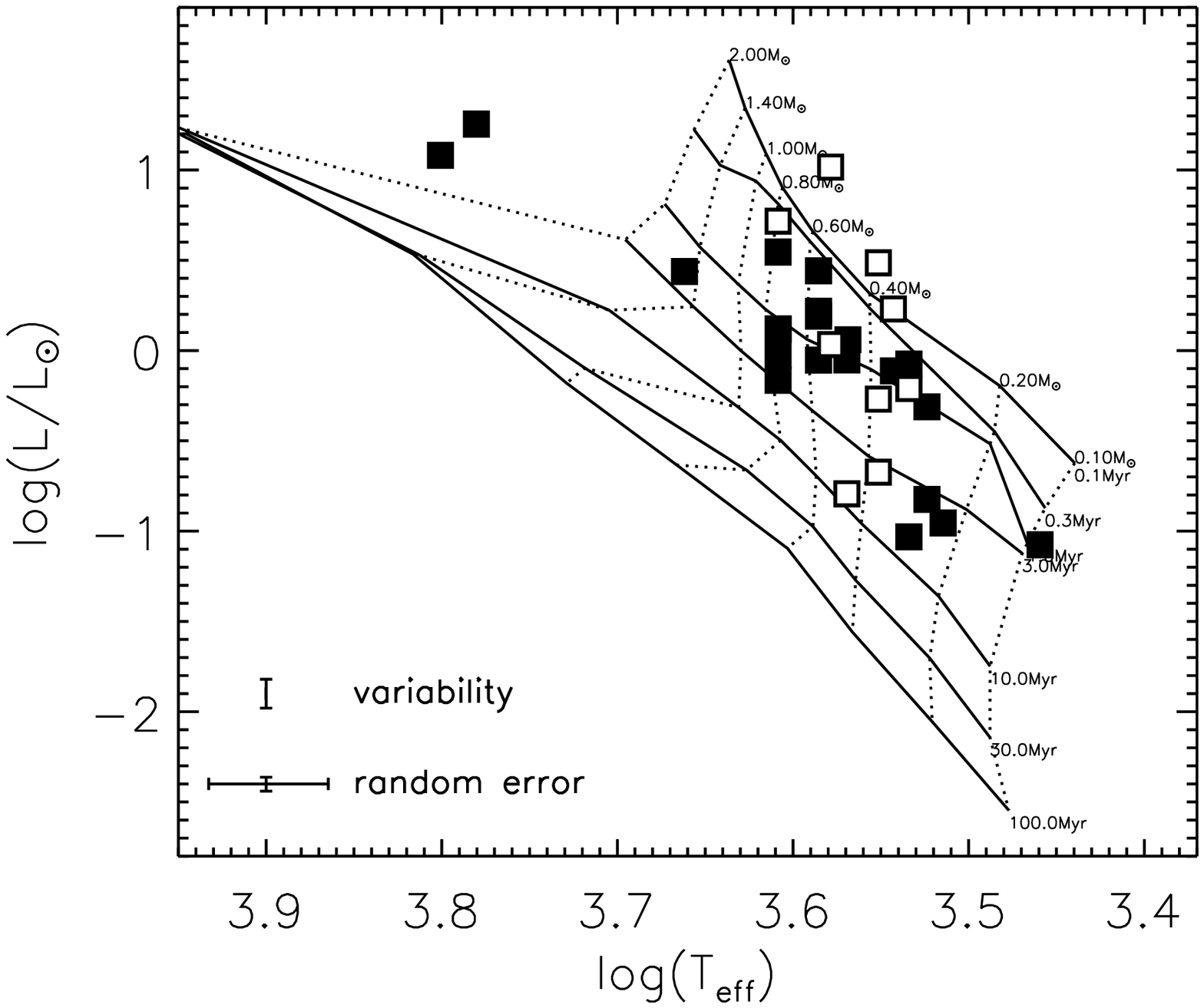}
  \caption{HR-diagram with evolutionary tracks from \citet{sie00}. Open symbols are target components with high veiling $r_K>0.2$, filled symbols have $r_K\le0.2$. Typical uncertainties from random errors and the resulting uncertainty from 0.2\,mag intrinsic photometric variability \citep{car01a} are shown in the lower left. For the determination of the parameters (age \& mass, see Table \ref{tab:componentparameters}), more tracks and isochrones were used than shown in the plot; they were omitted for a clearer illustration.}
  \label{fig:HRD}
\end{figure}

Effective temperatures and luminosities were used to derive ages and masses by comparison with evolutionary tracks from \citet{sie00}. Fig.~\ref{fig:HRD} shows the position on the HR-diagram of all components with measured $T_\mathrm{eff}$ and $L_*$.
The derived masses are -- except for three targets -- below 1\,$M_\odot$, and ages are found to be in the range of $10^4$--$10^7$\,yr, with an average of 0.97\,Myr.
One object, the secondary component of [HC2000]\,73, has an estimated mass of $0.09\pm0.05$\,$M_\odot$, indicating that it is a possible substellar object. 
The details of this binary are discussed in an accompanying paper \citep[\emph{in prep.}]{pet11}.

It is apparent that the targets that are classified as the youngest, are those with the highest veiling. Their high luminosities thus probably do not represent extreme youth but are rather an indication of hot circumstellar material contributing near-infrared flux even in $J$-band. This is not properly accounted for when extracting ages from the HR diagram because the infrared-excess in the $J$-band is unknown for our targets and therefore not subtracted from their brightness. A consequence is the apparent non-coevality of binary stars with at least one component with high veiling, as discussed in Sect.~\ref{sec:relativeExtinctions} and Fig.~\ref{fig:PrimaryAge-SecondaryAge}. Furthermore, apparently old ages can be caused by underestimated extinctions, making these targets appear underluminous and thus too old. However, since the evolutionary tracks for stars of a certain mass are almost vertical ($T_\mathrm{eff}\approx \mathrm{const}$) in this part of the HR diagram, the uncertainty in luminosity does not translate into an equally large uncertainty in mass.

\subsection{Accretion: $W($Br$\gamma)$, $L_\mathrm{acc}$, and $\dot{M}_\mathrm{acc}$}
The accretion activity of each target component was inferred from the Br$\gamma$ emission feature at 21665\,\AA. Our measurements are described in this section.
\subsubsection{Equivalent widths of Br$\gamma$ emission}\label{sec:WBr}
We measured the equivalent width of the Br$\gamma$ line
\begin{equation}\label{eq:WBr}
  W_{\mathrm{Br}\gamma} = \int_{\mathrm{Br}\gamma}\frac{F_\lambda-F_{\mathrm{c}}}{F_{\mathrm{c}}}d\lambda
\end{equation}
in an interval of width 56\,\AA\ around the center of Br$\gamma$ (which was individually fit and recentered between 21654\,\AA\ and 21674\,\AA) for all target components. The integration interval was chosen to ensure that we include as much line flux as possible, while minimizing the influence from the continuum noise around the line at the spectral resolution of our observations. The continuum $F_\mathrm{c}$ was determined from a linear fit to the local pseudo-continuum in a region of width $130$\,\AA\ shortward and longward of the integration limits. The above definition of equivalent width (eq.~\ref{eq:WBr}) returns positive values for emission lines and is negative in case of absorption.

To assess the significance of Br$\gamma$ emission, we inferred the noise level at the position of Br$\gamma$ from the surrounding continuum. This allowed us to derive a measure for the probability of the presence of a gaseous accretion disk in a binary component (see Appendix~\ref{sec:app:statistics}).

The procedure of measuring equivalent widths was performed in two stages. In the first stage, the method was applied to the uncorrected spectra to serve as a measure of accretion disks (Appendix~\ref{sec:app:statistics}), where we sought to assess the significance of the measured emission features and thus do not wish to introduce additional uncertainties by means of further modifications of the spectra. However, spectral features were reduced in strength when veiling was imposed on the spectrum and thus the Br$\gamma$ equivalent width had to be corrected for $r_K$ to represent the actual emission emitted from the accretion process. This required a second step of either calculating $W_\lambda = W_\lambda^\mathrm{measured}\cdot(1+r_\lambda)$, as inferred from eqs. (\ref{eq:F}) and (\ref{eq:WBr}), or applying eq.~(\ref{eq:F}) with $r_K$ values from Table~\ref{tab:componentparameters} to the reduced spectra (for which extinction does not change the equivalent width and can be assumed to be equal to zero for this calculation). Both options returned very similar results. We chose to modify the spectra and remeasure $W_{\mathrm{Br}\gamma}$ to also obtain good estimates for the continuum flux noise, which does not necessarily transform in the same way. The results for this measurement of the actual equivalent widths of the Br$\gamma$ emission are listed in Table~\ref{tab:componentparameters}. When $r_K$ was unknown, no correction was applied and values in Table~\ref{tab:componentparameters} correspond to lower limits of $W_{\mathrm{Br}\gamma}$.

\subsubsection{Br$\gamma$ equivalent widths versus NIR excess}\label{sec:brgammavsnirexcess}
\begin{figure}
  \centering
  \includegraphics[angle=0,width=0.48\textwidth]{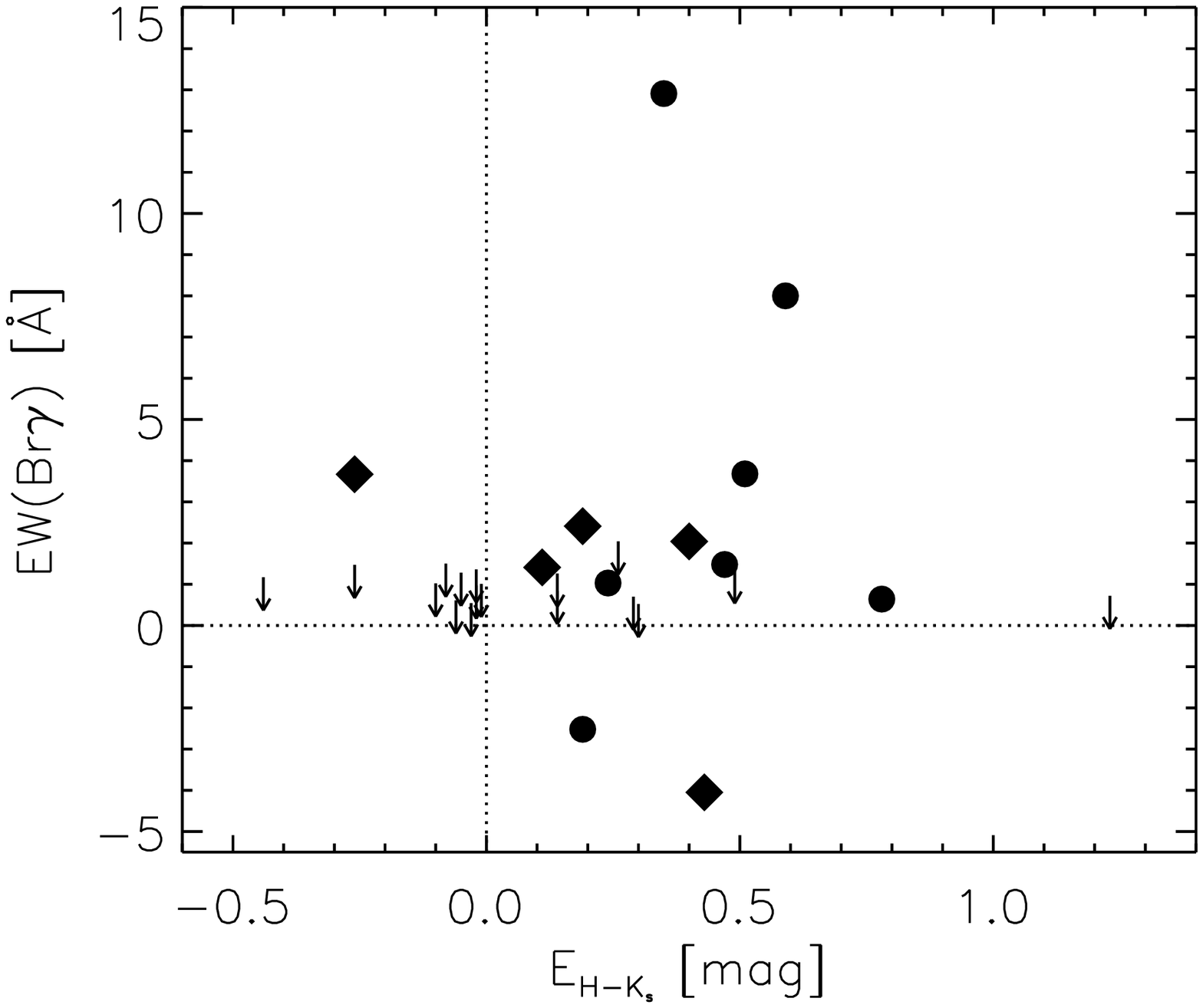}
  \caption{Veiling-corrected Br$\gamma$ equivalent width as a function of NIR excess in ($H$$-$$K_\mathrm{s}$) color. Circles represent primary, and diamonds secondary binary components. The detection limits of target components with insignificant equivalent widths are marked with arrows.}
  \label{fig:excessvsbrg}
\end{figure}

We compared the color excesses $E_{H-K_\mathrm{s}}$, which measure the existence of hot circumstellar material around each binary component \citep{cie05}, with our veiling-corrected Br$\gamma$ emission values. Fig.~\ref{fig:excessvsbrg} shows that, while all but one target with significant emission in Br$\gamma$ exhibit a NIR excess, the opposite is not true and many targets with NIR excess are found that do not show significant signs of hydrogen emission.
This imbalance is well-known and discussed further in Sect.~\ref{sec:accretiondiskfraction}.

The figure shows some interesting features. It seems that there is a slight systematic offset in the calculated excesses relative to the origin, as we see a number of targets clustering around $E_{H-K_\mathrm{s}}\approx-0.03$. Assuming that these are targets with no significant color excess, we can infer a small mismatch between the color scales of the theoretical and measured values used to derive $E_{H-K_\mathrm{s}}$. This is partly a consequence of using dwarfs for the theoretical colors $(H$$-$$K_\mathrm{s})_\mathrm{theor}$ instead of pre-main sequence stars. When using the colors of pre-main sequence stars derived by \citet{luh10}, we measured an average shift of 0.015\,mag towards redder $H$$-$$K_\mathrm{s}$ colors that moved the accumulation of low color-excess components towards the origin. Since this correction was small and the \citeauthor{luh10} spectral sequence does not cover all spectral types of our sample, we use color-excesses derived from the dwarf colors. In addition, the anticipated systematic offset was negligible compared to the possible photometric variability of $\sim$0.2 magnitudes \citep{car01a}. Although variability does not change our (qualititative) conclusions, it might be a reason for the position of the targets with negative excess with and without significant Br$\gamma$ emission.

Targets in the bottom right quadrant display Br$\gamma$ in absorption; since we also detect a NIR emission excess, there is a possibility that Br$\gamma$ emission generated by accretion cancels with part of the absorption feature. To estimate the real emission strength, one could measure and subtract the strength of Br$\gamma$ absorption from photospheric standards of the same spectral type. This concerns, however, only the components of JW\,260, a binary that was excluded from the discussion and conclusions owing to its earlier spectral type. We thus skip a more thorough evaluation of the accretion state of this binary here.

\subsubsection{Br$\gamma$ line luminosity and mass accretion rates}\label{sec:massaccretion}
For target components with full knowledge of extinction and veiling, we calculated Br$\gamma$ line luminosities and the mass accretion rate $\dot{M}_\mathrm{acc}$. The accretion luminosity was derived through
\begin{equation}\label{eq:Lacc}
  \log(L_\mathrm{acc}) = (1.26\pm0.19)\log(L_{\mathrm{Br}\gamma}/L_\odot)+(4.43\pm0.79)
\end{equation}
\citep{muz98} where the Br$\gamma$ line luminosity is defined as
\begin{equation}
  L_{\mathrm{Br}\gamma} = 4\pi r^2 \int_{\mathrm{Br}\gamma}(F_\lambda-F_{\mathrm{c}})\,d\lambda
\end{equation}
in the same integration limits used for the equivalent width and using a distance to the ONC of $r=414\pm7$\,pc \citep{men07}. To measure the luminosity of the Br$\gamma$ line, the spectra must be flux-calibrated. This was achieved by comparing the $K_\mathrm{s}$-band photometry from Table~\ref{tab:componentmagnitudes} with synthetic photometry obtained by convolving our measured spectra with the 2MASS $K_\mathrm{s}$ filter curve and integrating with $zp(F_\lambda) = 4.283\times 10^{-7}$ for the zeropoint\footnote{http://www.ipac.caltech.edu/2mass/releases/allsky/doc/sec6\_4a.html}. Since the $K_\mathrm{s}$ filter curve extends to slightly bluer wavelengths than our NACO spectra, we had to extrapolate the spectrum. To estimate the impact of our linear extrapolation on the integration result, we assumed several extrapolations with different slopes (up to unreasonable values). The resulting variation in $L_{\mathrm{Br}\gamma}$ is small, mainly because the extrapolated part coincides with the steep edge of the filter and was kept as an additional uncertainty. After correcting the calibrated spectra for veiling and extinction, the data was multiplied with the filter curve, integrated over $K_\mathrm{s}$-band, and converted to $L_\mathrm{acc}$ according to eq.~(\ref{eq:Lacc}). Uncertainties were estimated from error propagation of all involved parameters including the extrapolation error and the empirical uncertainties in eq.~\ref{eq:Lacc}. A source of additional uncertainty that we could not quantify in greater detail from our observations is variability. Since the photometry was not taken simultaneously with our spectral observations, the calibration of our spectra might suffer from additional uncertainty when these young targets were observed in different states of activity. Since we had no estimate of the size of this effect for an individual target, we did not introduce any correction but wish to caution that individual accretion luminosities might be offset from the true value. However, assuming that the effect of variability in the accretion luminosities is random, the sample statistics should not be biased. 

The mass accretion rate was calculated according to \citet{gul98} as
\begin{equation}
  \dot{M}_\mathrm{acc} = \frac{L_\mathrm{acc}\,R_*}{GM_*}\left(\frac{R_\mathrm{in}}{R_\mathrm{in}-R_*}\right)\quad,
\end{equation}
for a stellar radius $R_*$ and mass $M_*$ from Table~\ref{tab:componentparameters}, and the graviational constant $G$, assuming that material falls onto the star from the inner rim of the disk at $R_\mathrm{in}\approx5R_*$. The accretion luminosities and mass accretion rates are listed in Table~\ref{tab:componentparameters}.

\subsubsection{Is the Br$\gamma$ emission generated by magnotespheric accretion?}
The main source of Br$\gamma$ emission in \mbox{T\,Tauri} stars is often assumed to be magnetospheric accretion \citep[e.g.][]{bec10}. However, mechanisms such as stellar wind, disk wind, outflow, or photoevaporation of the disk by a nearby high-mass star can also contribute to the Br$\gamma$ emission observed in low-mass stars \citep{har90,har95,eis10}. While the emission region from magnetospheric accretion should be located close to the stellar surface, the sources of most other mechanisms are expected to be located further away at several stellar radii or even in the outer parts of the disk. Detecting a spatial displacement of the Br$\gamma$ emitting region from the star locus would refute the possibility of magnetospheric accretion as the origin of the emission and render other explanations more likely.

To test this, we used the sky-subtracted raw frames of the spectral observations, showing the spectral traces of both components of each target binary. The orientation of the images is such that the dispersion direction is roughly aligned along the columns of the detector, while the spatial information is aligned along the rows. In each row, we fit two Gaussians to the two profiles of the component spectra, thus measuring the spatial location and width of the spectral profile of the targets in each wavelength bin of size $\sim$5\,\AA. Depending on the S/N of the individual observations, the locations of the trace center can be determined to an accuracy of 0.008--0.10 pixels, which, at the spatial resolution of the observations of $\sim$0.027\,arcsec/pix, corresponds to $\sim$0.1--1.1\,AU at the distance of the ONC.

We found no significant offset at $\lambda$(Br$\gamma$) from the rest of the trace for any of the targets in which we observed Br$\gamma$ in emission. Neither did we detect extended emission in excess of the width of the spectral trace at similar wavelengths. This indicates that the emission indeed originates from a small region close to the stellar surface and is likely to be the product of magnetospheric accretion. However, extended or displaced emission may well be generated mainly perpendicular to the slit. In this case, a displacement of the emission peak could be observed in the dispersion direction. Since we did not measure frequency shifts of spectral features, we cannot exclude this possibility for any individual target component.

While the Br$\gamma$ emission seems to come from a region close to the star for all targets, we observed one target ({TCC\,55}) in which emission at $\lambda($H$_2) = 21218$\,\AA\ comes from an extended region around the star, rather than the star itself. We observed at least three diffuse H$_2$ emitting regions along the slit, each several pixels wide, one of them apparently surrounding the binary. We used K-band images from the MAD instrument \citep{pet08} to investigate the surrounding area for possible emission sources close to this target located only 0\farcm49 from $\theta^1$\,Ori\,C. We identified bow-shocks coinciding with the location of the H$_2$ features in the spectra. The diffuse emission around the binary itself might represent the remaining material of a proto-stellar envelope, which would be indicative of a young evolutionary state (i.e.\ Class\,I) of the binary.

\section{Discussion}\label{sec:discussion}

\subsection{Stellar parameters and sample biases}
To detect possible biases in the numbers derived from the components of our binary sample, we discuss the degree to which the populations of primaries and secondaries differ and whether they are typical of the ONC population.

\subsubsection{Spectral types}
\begin{figure}
  \centering
  \includegraphics[angle=0,width=0.48\textwidth]{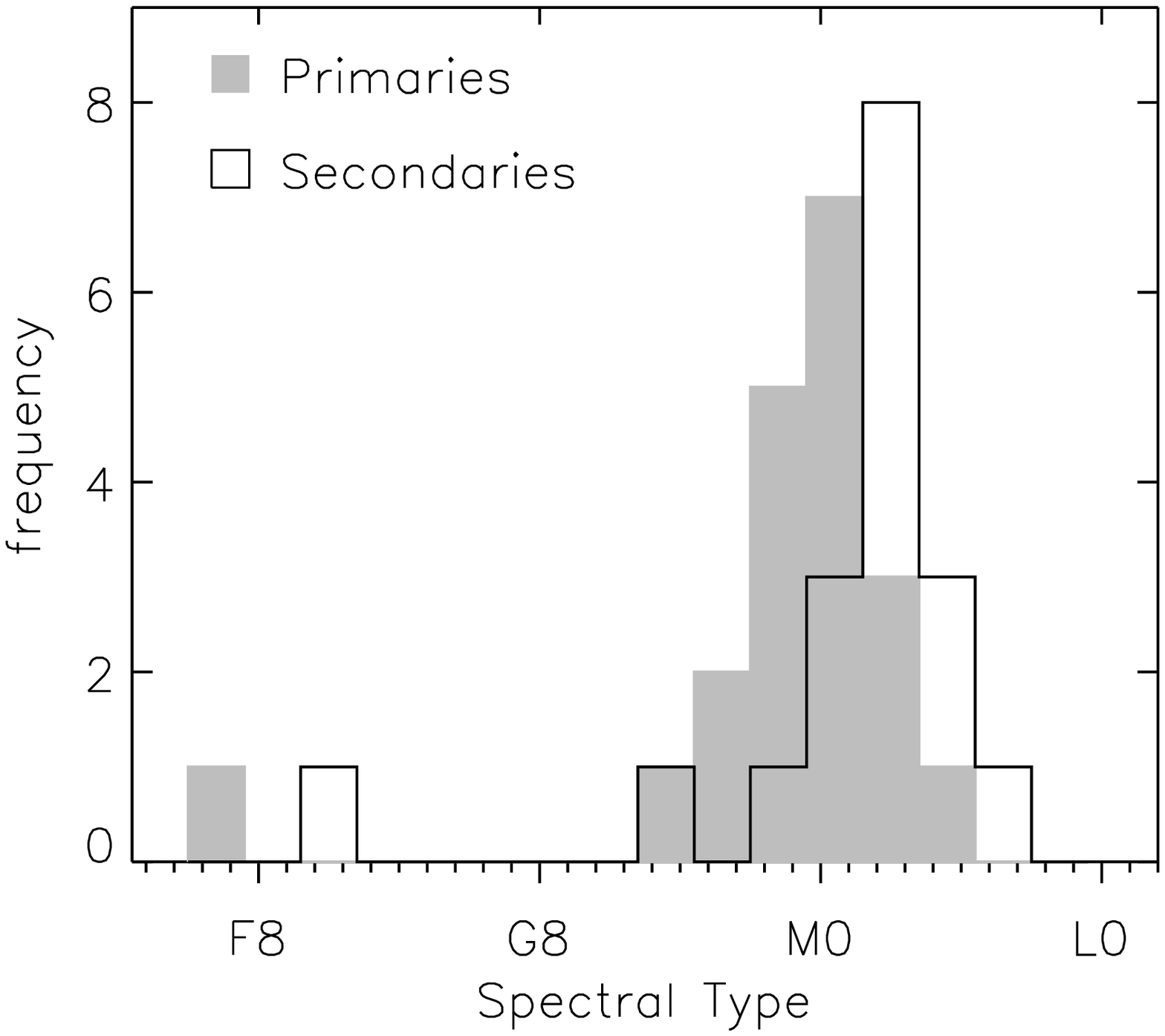}
  \caption{Number of target components per spectral type. Gray areas: primaries. Black outline: secondaries. The distributions of primaries and secondaries are only slightly different in spectral types.}
  \label{fig:spthistogram}
\end{figure}

It is known that the strengths of accretion parameters correlate with the mass of a star, which (considering the limited spread in age) can be represented by the typically more tightly constrained spectral type\footnote{That accretion signatures do correlate with spectral type can be seen from e.g. Table 7 in \citet{whi01}.}. In Fig.~\ref{fig:spthistogram}, we see that the spectral type distribution of the primaries peaks at slightly earlier spectral types of M0--M1 than the secondaries (M2--M3).
This difference is significant at 98\%, according to a Kolmogorov-Smirnov (K-S) test. However, both distributions are indistinguishable from the spectral types of the entire ONC population \citep{hil97}, with K-S probabilities for different distributions of 65\% and 57\% for primaries and secondaries, respectively. This means that both primaries and secondaries are 'typical' members of the ONC, whereas the primary and secondary spectral type distributions deviate slightly and differences in the derived parameters (such as accretion rates and Br$\gamma$-emission strength) can partly be attributed to the -- on average -- earlier spectral types of the primaries.

\subsubsection{Relative extinctions and ages}\label{sec:relativeExtinctions}
Interstellar extinction through embedding in the Orion molecular cloud expresses itself as a spatially variable source of extinction that is, nevertheless, very similar for all components of a stellar multiple. An additional source of extinction that can be very different even for components of the same binary can be caused by circumstellar material such as a (nearly edge-on) circumstellar disk, obscuration by the other component's disk, or a remaining dust envelope.

\begin{figure}
  \centering
  \includegraphics[angle=0,width=0.48\textwidth]{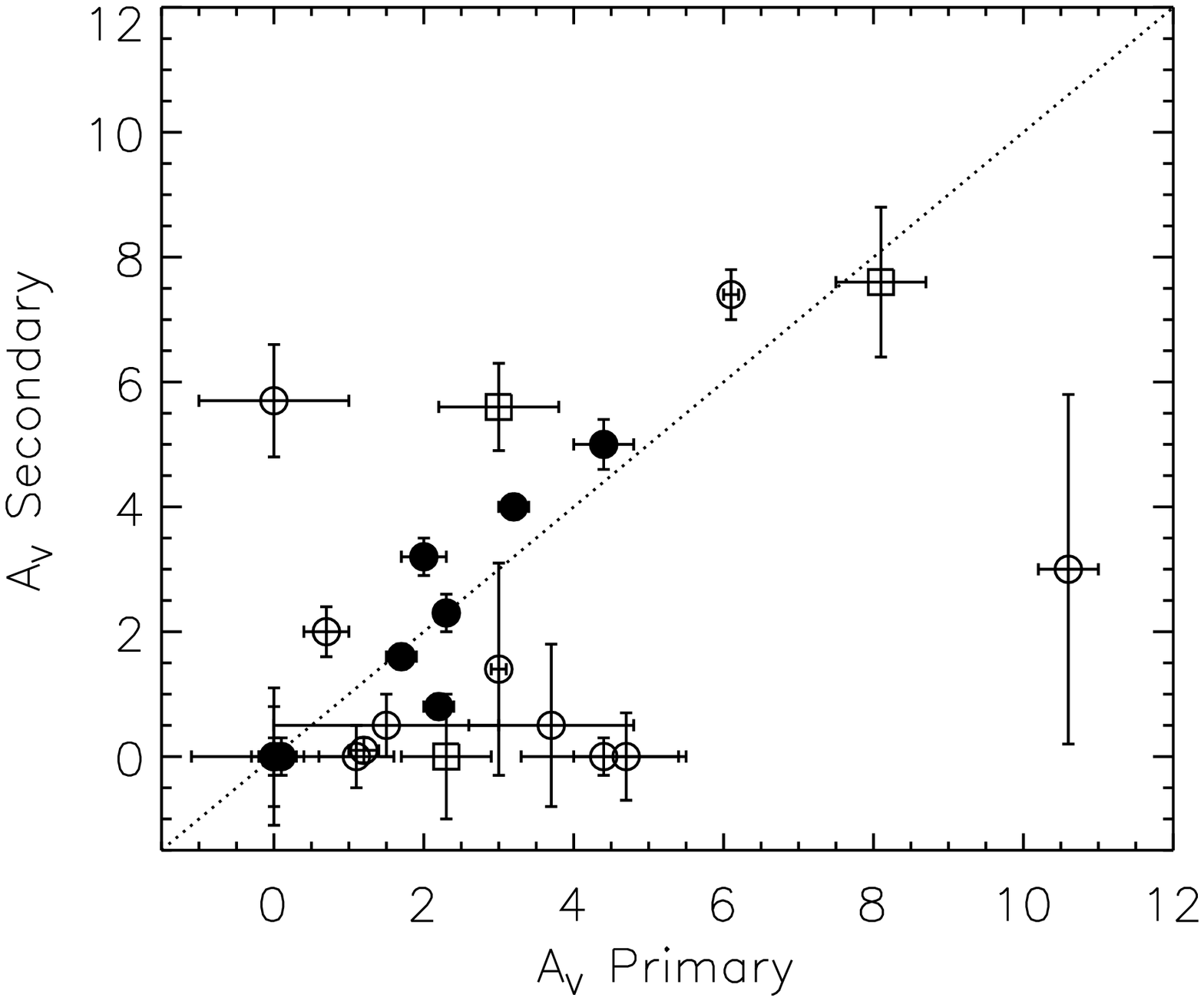}
  \caption{Extinction of the primary versus versus secondary components. Circles show targets with $A_\mathrm{V}$ determined from the dereddening of the CTT locus. Squares show the spectroscopically determined extinctions of {JW\,681}, {JW\,876}, and {JW\,959}, because photometric extinctions could not be measured. Filled symbols indicate targets with both components having a low veiling $r_K<0.2$. The dashed line corresponds to equal extinctions. }
  \label{fig:AVprimary-AVsecondary}
\end{figure}

In Fig.~\ref{fig:AVprimary-AVsecondary}, we can identify and verify the impact of the different sources of extinction.
Binaries composed of two low-veiling components have similar extinctions, according to the level of embedding in the cloud, whereas binaries with high-veiling components do not display such correlation. This might be due to dust material from a disk that is detected by means of its hot-continuum emission from magnetospheric accretion, i.e.\ veiling \citep[e.g.][]{bou07}. In particular, the magnitude of extinction is a strong function of the angle under which the system is observed with close to edge-on disks causing a strong reddening of the NIR colors \citep{rob06}. However, our observations do not enable us to determine to the inclinations of the circumstellar disks, hence conclusions about disk orientation and alignment cannot be drawn.

\begin{figure}
  \centering
  \includegraphics[angle=0,width=0.48\textwidth]{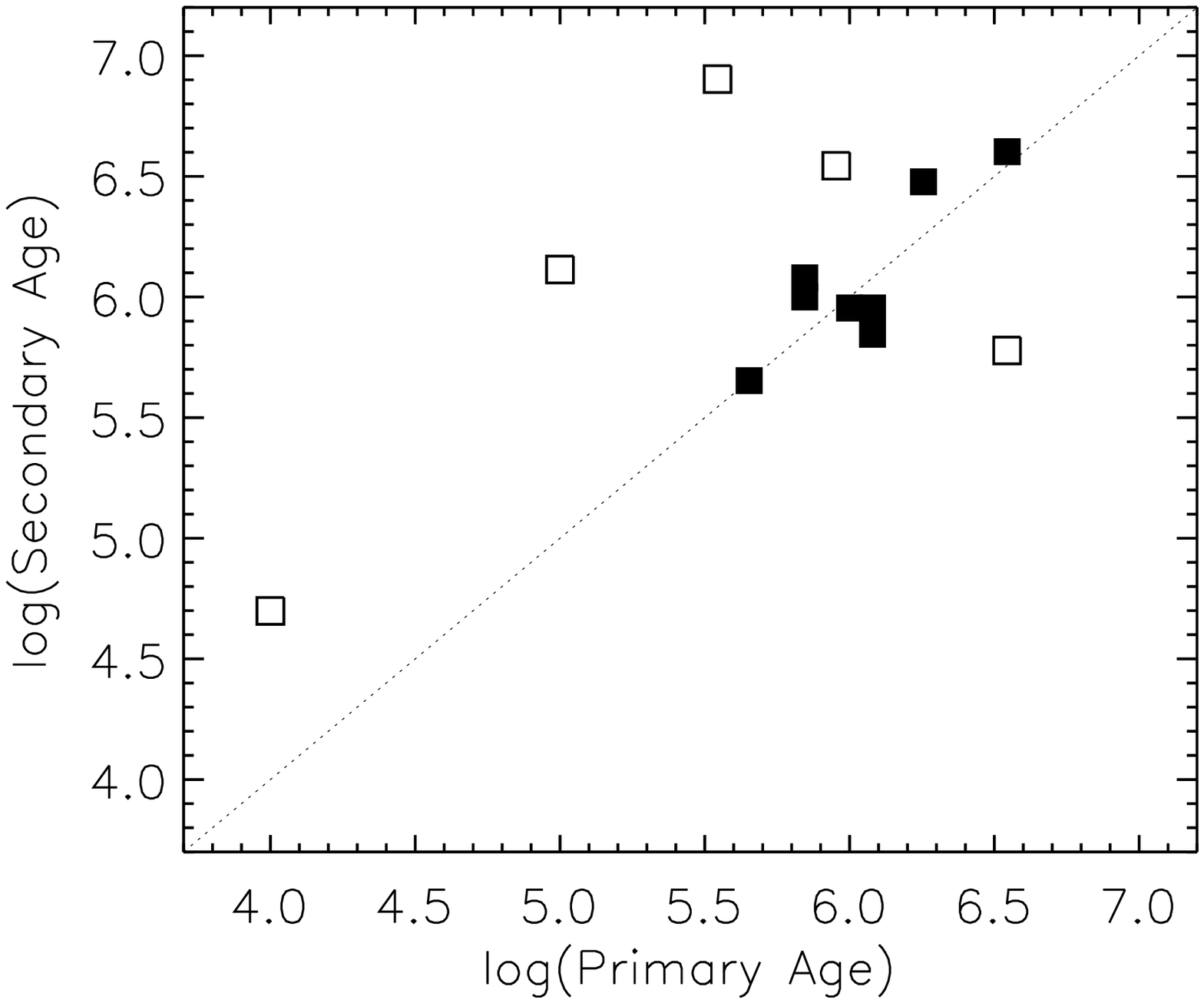}
  \caption{Relative ages for the primary and secondary components of all target binaries where both components could be placed in the HR diagram. Filled squares show targets where both components exhibit low veiling $r_K<0.2$. Targets with open squares have at least one component with strong veiling $r_K\ge0.2$, thus it is likely that the age estimation from the HR diagram is biased, since extra luminosity from accretion makes the targets appear brighter and thus younger.}
  \label{fig:PrimaryAge-SecondaryAge}
\end{figure}

The stellar components within a binary are close to being coeval. Fig.~\ref{fig:PrimaryAge-SecondaryAge} shows a well-defined correlation of primary and secondary ages for those binaries with little or no veiling.
The five binaries with at least one component of $r_K\ge0.2$ are clearly located off the sequence of coeval binaries, which is probably due to a non-negligible amount of veiling in the $J$-band for targets with high $r_K$ values. In our present study, we did not attempt to derive accurate absolute ages for our sample, but testing for equal ages within binaries does serve two purposes: \emph{i)} A sanity check. Assuming that binary components do form reasonably close in time \citep{kra09}, we confirmed that our derived $T_\mathrm{eff}$ and $L_*$ were accurately determined because they result in reasonable values when placed in an HR-diagram. We inferred that the derived parameters (e.g.\ $L_\mathrm{acc}$, $\dot{M}_\mathrm{acc}$) are of sufficient quality to help us derive the conclusions of this paper. \emph{ii)} The derived parameters have no age dependence. Since there is no systematic difference between the primary and secondary age of the binaries, we were able to exclude any dependence on age of the derived relative parameters between the populations of the primaries and secondaries. \emph{iii)} Coevality is consistent with the physical binarity.

While we derived ages $\tau=\log(\mathrm{age})$ that span a range of $\sim\!5.5\le\tau\le\sim\!6.5$ even for the well-behaved class of low-$r_K$ binaries, this can probably not be attributed to a real age spread. \citet{jef11} observed no difference in age between stars with and without disks in the ONC, indicating that the age spread must be shorter than the disk lifetime, i.e.\ significantly more confined than traditionally assumed. We conclude that not only observational uncertainties, but also an intrinsic scatter in the luminosities at constant age must be present. Despite the importance of this observation for the absolute ages of members of star forming regions, we assumed that our \emph{relative} age measurements apply, since the boundary conditions are very similar for both components of the same binary.

\subsection{Disk evolution around the components of visual ONC binaries}
The absolute and differential abundances of disk signatures in the components of our target binaries were found to be a function of binary parameters, as we discuss in the following sections.
\subsubsection{The disk fraction of binary components}\label{sec:accretiondiskfraction}
We derived the fraction of ONC binary components harboring an accretion disk. This number was compared to the single-star disk frequency in the ONC and to binary samples of other star forming regions to expose the effect of binarity and cluster environment on the evolution of circumstellar disks.

The probability density function of the disk frequency was derived in an Bayesian approach as well as the probability of an individual binary component harboring an accretion disk (see Appendix~\ref{sec:app:statistics}; individual disk probabilities in Table~\ref{tab:componentparameters}). As the strength of disk signatures depends on spectral type \citep[e.g.][]{whi01}, we excluded the binary {JW\,260} from further evaluation of disk frequencies, since its spectral type is considerably earlier than for the rest of the sample. We measured an accretion disk fraction of $F=35_{-8}^{+9}$\% in our sample of 42 spectroscopically observed binary components (Appendix~\ref{sec:app:statistics}). 

Analyses of the Br$\gamma$ line are known to return a smaller fraction of accretors than the H$\alpha$ emission feature, which is typically used to decide whether an individual star can be classified as a classical \mbox{T\,Tauri} star. \citet{fol01} observed a number of classical \mbox{T\,Tauri} stars in Taurus-Auriga, measuring NIR emission-line strengths including Brackett-$\gamma$ signatures. Twenty-four of their targets are in the range of K3--M6 spectral types, comparable to our sample, and three show no emission in Br$\gamma$. Hence, a fraction of $f=0.125$ of all classical \mbox{T\,Tauri} targets would have to have been misclassified as weak-line \mbox{T\,Tauri} stars from their Br$\gamma$ emission. We thus expected a number of $f/(1-f)\times (F\cdot N)\approx2$ (with $N=42$ target components) classical \mbox{T\,Tauri} stars to be classified as non-accreting. The corrected fraction of classical \mbox{T\,Tauri}s among binary star components in the ONC was thus $F_\mathrm{CTT}=40_{-9}^{+10}$\%. Since many studies refer to the fraction of H$\alpha$-detected classical \mbox{T\,Tauri} stars rather than the number of accretors from Br$\gamma$ emission, we used the latter number to compare it with studies of accretion disk frequencies.

\citet{hil98} and \citet{frs08} found a frequency of accretion disks bearing single stars in the ONC of 50\% and 55\%, respectively. Both are marginally (1$\sigma$ and 1.5$\sigma$) larger than our measured fraction of $40_{-9}^{+10}$\%. While the \citeauthor{hil98} sample was derived from the $I_\mathrm{C}-K$ color instead of $H\alpha$ measurements, the \citeauthor{frs08} sample is biased towards classical \mbox{T\,Tauri} stars making their estimate an upper limit of disk frequency. Both of these findings prevent us from drawing firm conclusions about the difference between the disk frequencies of single stars and binaries. This evidence of a lower disk frequency around 100--400\,AU binary components will hence need future confirmation from observations using comparable diagnostics (preferably Br$\gamma$) in an unbiased comparison sample.

Similarly, we found evidence of an underrepresentation of dust disks in Orion binaries. Sixteen out of 27 target components (excluding JW\,260) with measured $H$$-$$K_\mathrm{s}$ colors show signs of a dust excess, that is a fraction of 59$\pm$15\% compared to the values of 55\%--90\% found for single stars in the ONC by \citet{hil98} and 80$\pm$8\% from \citet{lad03}. Since these numbers were derived using indicators other than the $H$$-$$K_\mathrm{s}$ excess, which is known to typically return a comparably small fraction of dust disks compared to e.g.\ $K$$-$$L$ \citep{hil05}, this evidence cannot be quantified in greater detail.  

Our numbers suggest that there is a higher frequency of target components with inner dust disks (59$\pm$15\%) than accretion disks ($40_{-9}^{+10}$\%). This discrepancy agrees with observations of single stars in various star-forming regions, where \citet{fed10} concluded that accretion disks decay more rapidly than dust disks in a particular cluster.

The presence of dust and accretion disks around the binary components of our sample is thus consistent with \emph{i)} observations of single stars in a variety of young clusters that dust disks are more abundant than accretion disks in the same cluster and \emph{ii)} the expectation that disk lifetimes are shorter for disks in binary systems than for single stars with comparable properties. The latter is also theoretically motivated by the missing outer disk through dynamical truncation \citep{art94} and the resulting reduced feeding of the inner disk from the outer disk material \citep{mon07}. As expected, disk frequency is a function of binary separation (also see Sect.~\ref{sec:differentialdiskevolution}). For example, \citet{cie09} found the component disk frequency in tight $<$40\,AU binaries of Taurus-Auriga to be significantly lower, at less than one half of the single-star disk frequency. Both their wider binaries (40--400\,AU) and our sample (100--400\,AU), however, have disk frequencies lower but comparable to single stars, indicating that there is a separation dependent mechanism.

\subsubsection{Synchronized disk evolution in ONC binaries}\label{sec:synchronizedevolution}
\begin{figure}
  \centering
  \includegraphics[angle=0,width=0.48\textwidth]{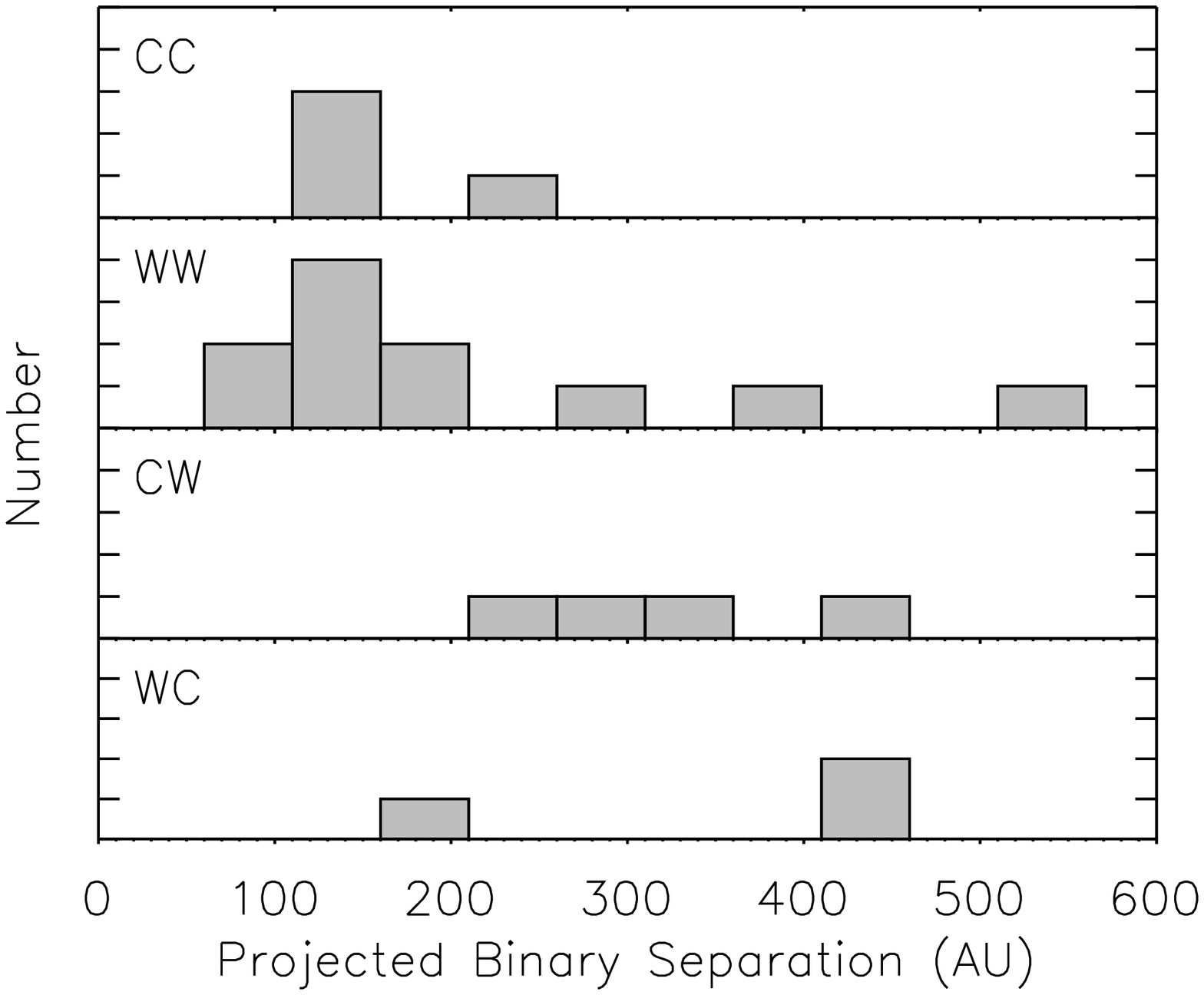}
  \caption{Histograms of binary separation as a function of component accretion-type (the y-axis tickmarks indicate one binary each). Most of the close binaries are of type WW or CC, i.e.\ synchronized in their disk evolutionary state. }
  \label{fig:sephistogram}
\end{figure}

We detected a significant overabundance of close, $\lesssim$200\,AU pairs of equal emission state, i.e.\ with both components accreting (CC)\footnote{To ease the reading of the binary categories, we use the common abbreviations 'C' for accreting components (referring to \emph{c\/}lassical \mbox{T\,Tauri} Stars), and 'W' for non-accreting components (\emph{w\/}eak-line \mbox{T\,Tauri} stars) where in the designation of a binary pair (e.g.\ 'CW') the first position describes the primary (here 'C') and the second the secondary (here 'W') component state. We note, however, that we only refer to the presence of accretion as measured through Br$\gamma$, which is correlated with but not equal to the the distinction between weak-line and classical \mbox{T\,Tauri} stars (see also the discussion Sect.~\ref{sec:accretiondiskfraction}).} or both components showing no accretion (WW) signatures. This is apparent in Fig.~\ref{fig:sephistogram} and a K-S test indicates 99.5\% probability that the separation distribution of equal pairs (CC and WW) differs from the separations of mixed pairs (CW and WC).

To investigate a possible correlation between the synchrony of disks in ONC binaries and their separations, we split our sample into binaries with projected separations larger and smaller than 200\,AU. From the content of accreting and non-accreting components in the two separation bins, we predicted the average number of CC, WW, and mixed systems by random pairing, and compare it to our measured distribution. If there were no correlation between the evolution of both components of the same binary, the randomly paired sample of the same number of components (W and C) should be consistent with our observed sample. With 18 non-accreting components and 6 accretors in the sample of 12 binaries with separations $<$200\,AU, we expect an average of $\sim$1.2$\times$CC, $\sim$6.9$\times$WW, and $\sim$4.1 mixed systems, as predicted by a Monte Carlo simulation. However, we found 3$\pm$1$\times$CC, 9$\pm$0$\times$WW, and 0$^{+2}_{-0}$ mixed pairs\footnote{uncertainties are derived from the possibility of a binary changing classificiation within its 1$\sigma$ limit of W$_{\mathrm{Br}\gamma}$, i.e.\ a CC might turn into a CW, if its secondary is classified as C but with an equivalent width that is less than 1$\sigma$ away from W$_\mathrm{min}$.}. This is clearly incompatible with the prediction of random pairing. On the contrary, wide-separation binaries are well described through random pairing with predicted values of 2.2$\times$CC, 4.1$\times$WW and 4.8$\times$mixed and measured values of 1$^{+1}_{-0}\times$CC, 3$^{+2}_{-1}\times$WW and 7$^{+1}_{-2}\times$mixed.

\citet{whi01} observed a similar underdensity of mixed pairs among binaries with separations $<$210\,AU in a sample of 46 binaries in the Taurus-Auriga star-forming association. They concluded that synchronized evolution, which they attributed to the existence of a circumbinary reservoir, can more or less equally replenish the circumprimary and circumsecondary disks, as previously suggested by \citet{pra97}. 

Can circumbinary disks of sufficient size survive in the ONC and thus be the cause for the synchronization of disk evolution? The typical size of a disk in the Trapezium region is below 200\,AU \citep{vic05} and only 3 of 149 analyzed systems -- the authors claim completeness for large disks $>$150\,AU at moderate extinctions -- were found to have disk sizes $>$400\,AU. However, circumbinary disks have inner radii of at least twice the binary separation \citep{art94}. This requires circumbinary disk sizes of more than 400\,AU in diameter for a binary with 100\,AU separation and even $\gtrsim$800\,AU circumbinary disks for 200\,AU binaries. Assuming that dynamical interactions \citep{olc06} and photoevaporation \citep{man09a} are the reason for disk truncation in the ONC, the observed size limits of single star disks should also apply to circumbinary matter. This implies that circumbinary disks of binary systems $>$100\,AU should not be largely abundant since they are typically truncated to radii below the dynamically induced inner hole radius. These considerations render it unlikely that stable circumbinary disks are the reason for disk-synchronization in the inner regions of the Orion Nebula cluster.

\begin{figure}
  \centering
  \includegraphics[angle=0,width=0.48\textwidth]{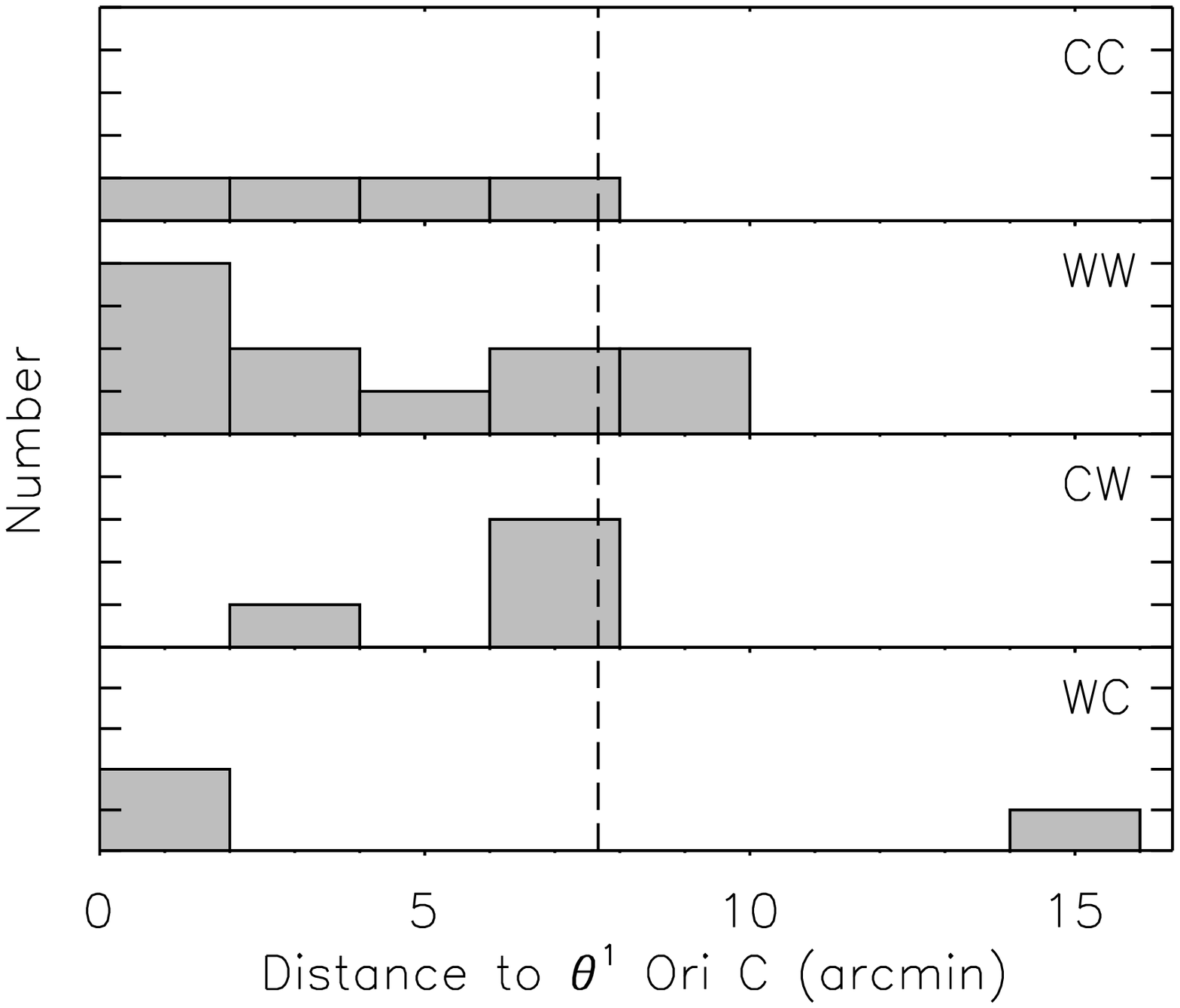}
  \caption{Histograms of distance to $\theta^1$\,Ori\,C as a function of accretion type. The histogram shows no indication that binaries of any type are more likely at any distance. The vertical dashed line shows the radius \citep[460\arcsec;][]{rei07} inside which the ratio of wide binaries (0\farcs5--1\farcs5) to close binaries (0\farcs15--0\farcs5) drops considerably, probably owing to dynamical interaction.}
  \label{fig:disttocenter}
\end{figure}

In agreement with this, none of these systems that were observed with circumbinary material (e.g.\ GG Tau, \citealt{dut94}; V892\,Tau, \citealt{mon08}; Orion proplyd 124-132, \citealt{rob08}) have binary separations of more than 100\,AU. This points to the possible universality of the result. \citet{vic05} observed no trend in the disk sizes with distance to $\theta^1$\,Ori\,C out to $\sim$4\arcmin, which means that Orion disks are small, independent of their position in the inner $\sim$50\,arcmin$^2$ of the cluster. Furthermore, we would expect to see a larger ratio of CC binaries to WW binaries at large distances from $\theta^1$\,Ori\,C when circumbinary disks are present only in the outer parts of the cluster and transfer a significant amount of material to the individual stellar disks. The distribution of CC and WW in our ONC does not, however, increase with distance to $\theta^1$\,Ori\,C (Fig.~\ref{fig:disttocenter}). 
Finally, although in Taurus \citep{and07} more disks with large radii have been observed than in the ONC, probably owing to its weaker dynamical interactions and irradiation, the similarity of the observed parameters (such as the 200\,AU limit for synchronization) seems to suggest that the disk feeding mechanism in Taurus binaries is probably similar to that in Orion, i.e.\ not due to replenishment from circumbinary disks.

\begin{figure}
  \centering
  \includegraphics[angle=0,width=0.48\textwidth]{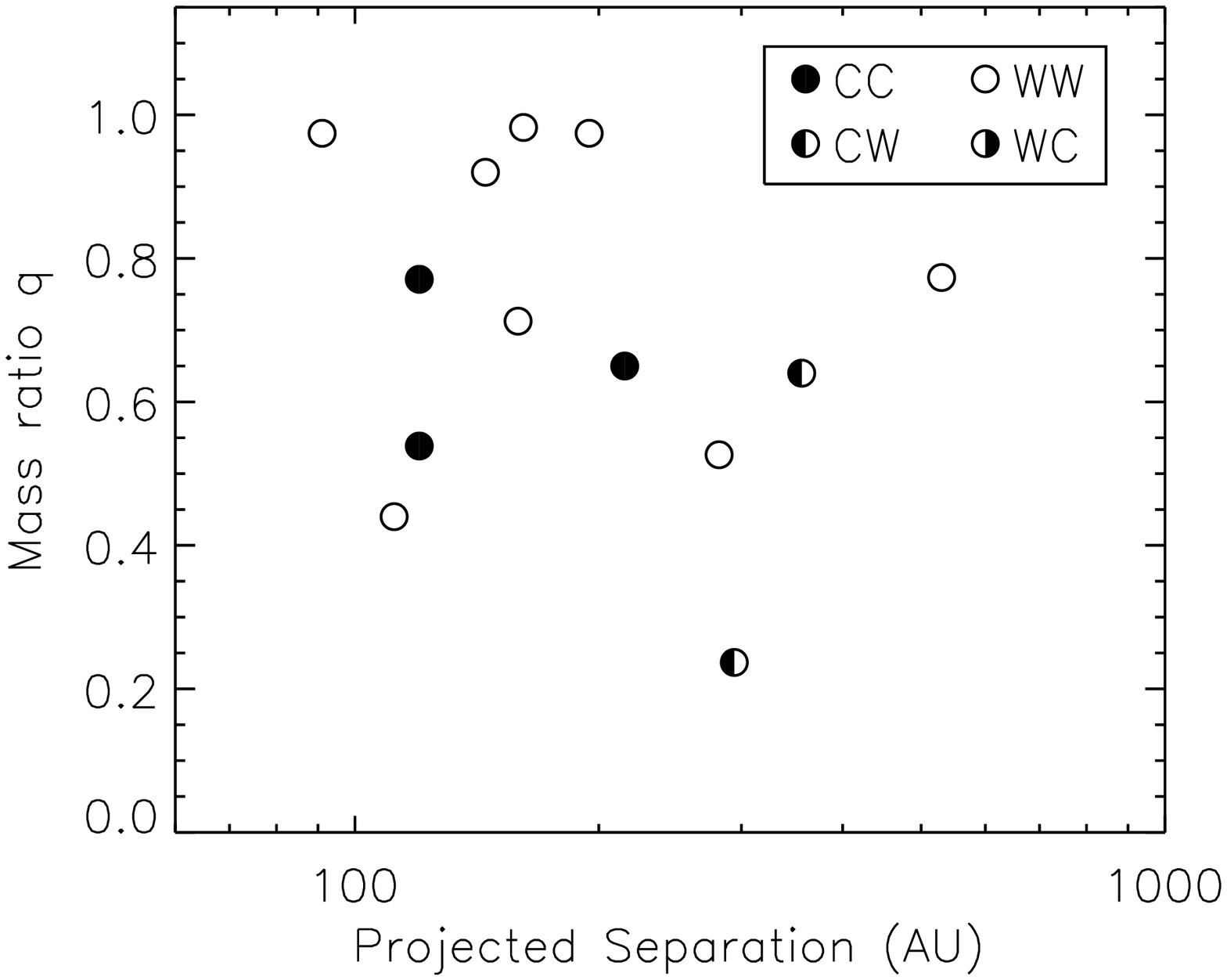}
  \caption{Mass ratios of binaries as a function of projected binary separation, indicating whether both binary components are accreting (filled circles, CC), neither component is accreting (open circle, WW), or either component shows signs of accretion (half-open circles = CW and WC). This plot only contains 13 targets, since masses could not be derived for all binary components in the sample. For some target binaries with mass ratios close to 1, the primary (as estimated from the NIR colors and listed in Table~\ref{tab:componentmagnitudes}) turned out to be the less massive component. The mass ratios of those targets were calculated as the inverse leading to $q$-values $\lesssim$1.}
  \label{fig:sep-q}
\end{figure}

What other mechanism could synchronize the circumprimary and circumsecondary disks in $\lesssim$200\,AU systems? Since mass accretion rates are a function of stellar mass \citep{whi01} and disk truncation radii are similar in equal mass systems \citep{art94}, synchronization of disk evolution might arise, if close binary systems are preferentially equal mass systems. For 13 binaries of the sample, we were able to derive masses of both binary components (see Fig.~\ref{fig:sep-q}).
For these, we observe that all (four) systems with mass ratios of q$>$0.8 and separations $<$200\,AU are of WW type, agreeing with the hypothesis of high mass ratios being the cause of synchronized disk evolution. The significance of this result, however, is low. There is a 21\% chance that the mass ratios of close ($<$200\,AU) systems are drawn from the same parent distribution as mass ratios of wider pairs (K-S test). A larger sample of spatially resolved spectroscopic observations of pre-main sequence binaries is needed to decide whether mass ratios are the main driver of the synchronization of binaries closer than 200\,AU.

\subsubsection{Differential disk evolution in binaries}\label{sec:differentialdiskevolution}
Mixed pairs with accreting (CW) and non-accreting primaries (WC) -- within the uncertainties -- are equally abundant: CW pairs appear 4$\pm$2 times while WC are measured 3$\pm$1 times. Although not statistically significant, this is evidence against a strong preference for primaries to have longer lived disks than the less massive secondary. However, longer lived disks around primaries are suggested by theory since disks around secondaries are truncated to smaller radii \citep{art94} and dissipation times are predicted to scale like $R^{2-a}$ with $R$ the disk radius and $a$$\approx$1--1.5 \citep[and references therein]{mon07}. \citet{mon07} found that their measured overabundance of 14$\times$CW versus 6$\times$WC is consistent with this effect taking place, however, with other factors (i.e.\ initial disk conditions) having a more a dominant impact on the lifetimes of circumprimary and circumsecondary disks than the differential scaling with $R$, which is only strong for binaries with low mass ratios $q$$\le$0.5. Our data agree with this proposed \emph{weak} correlation of the binary mass ratios with the abundance of CW-binaries in Orion, though this result is limited in significance by the small number of mixed systems in our sample.

It is noteworthy that the existence of mixed pairs, together with their property of having wider separations (Fig.~\ref{fig:sephistogram}), can introduce difficulties in the interpretation of binary studies that do not resolve their targets into separate components. \citet{cie09} used NIR photometry of unresolved binaries with known separations from several star-forming regions (Taurus, $\rho$-Oph, Cha\,I, and Corona Australis) to infer a smaller separation in binaries with no accreting components than in accreting binaries. Besides the proposed shorter disk lifetimes around close binary components, there is an alternative interpretation of their data that they did not discuss. Since they did not resolve binaries into separate components, they were unable to distinguish CC, CW, and WC-type binaries, but merged them all into the category of having at least one disk. As we showed earlier, however, the separation distribution of CW and WC binaries differ significantly from equal-accretion binaries including CC (Fig.~\ref{fig:sephistogram}). When joining the three categories with at least one accreting component, the combined separations display a distribution with on average larger separations than the WW distribution. Only from our resolved population, can we interpret this as a lack of mixed (CW+WC) pairs and not a lack of accreting (CW+WC+\emph{CC}) components in close binaries.

\subsubsection{Accretion luminosities and mass accretion rates}\label{sec:massaccretionrates}

\begin{figure}
  \centering
  \includegraphics[angle=0,width=0.48\textwidth]{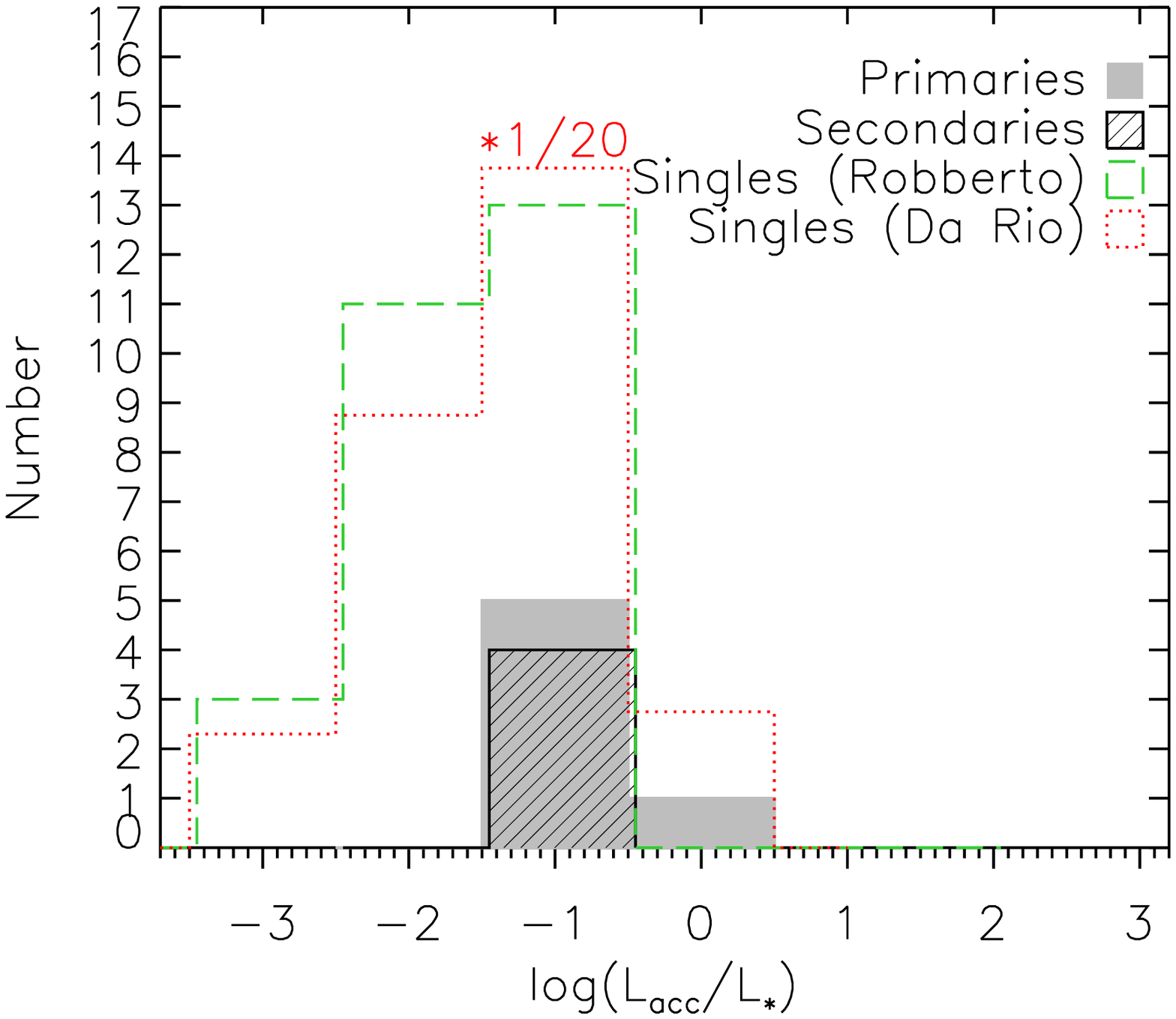}
  \caption{Histogram of the accretion luminosities of the primaries (gray shaded area) and secondaries (hatched) calculated from the Br$\gamma$ emission. For comparison, the accretion luminosity distributions of single stars of Orion from \citet[dashed outline]{rob04} and \citet[dotted outline, scaled by a factor of 1/20 for clear comparability]{da_10} are overplotted, both limited to the same range of stellar masses as in our binary survey.
The distributions are slightly offset relative to each other to make them more visible.}
  \label{fig:accretionluminosityHistogram}
\end{figure}

\begin{figure}
  \centering
  \includegraphics[angle=0,width=0.48\textwidth]{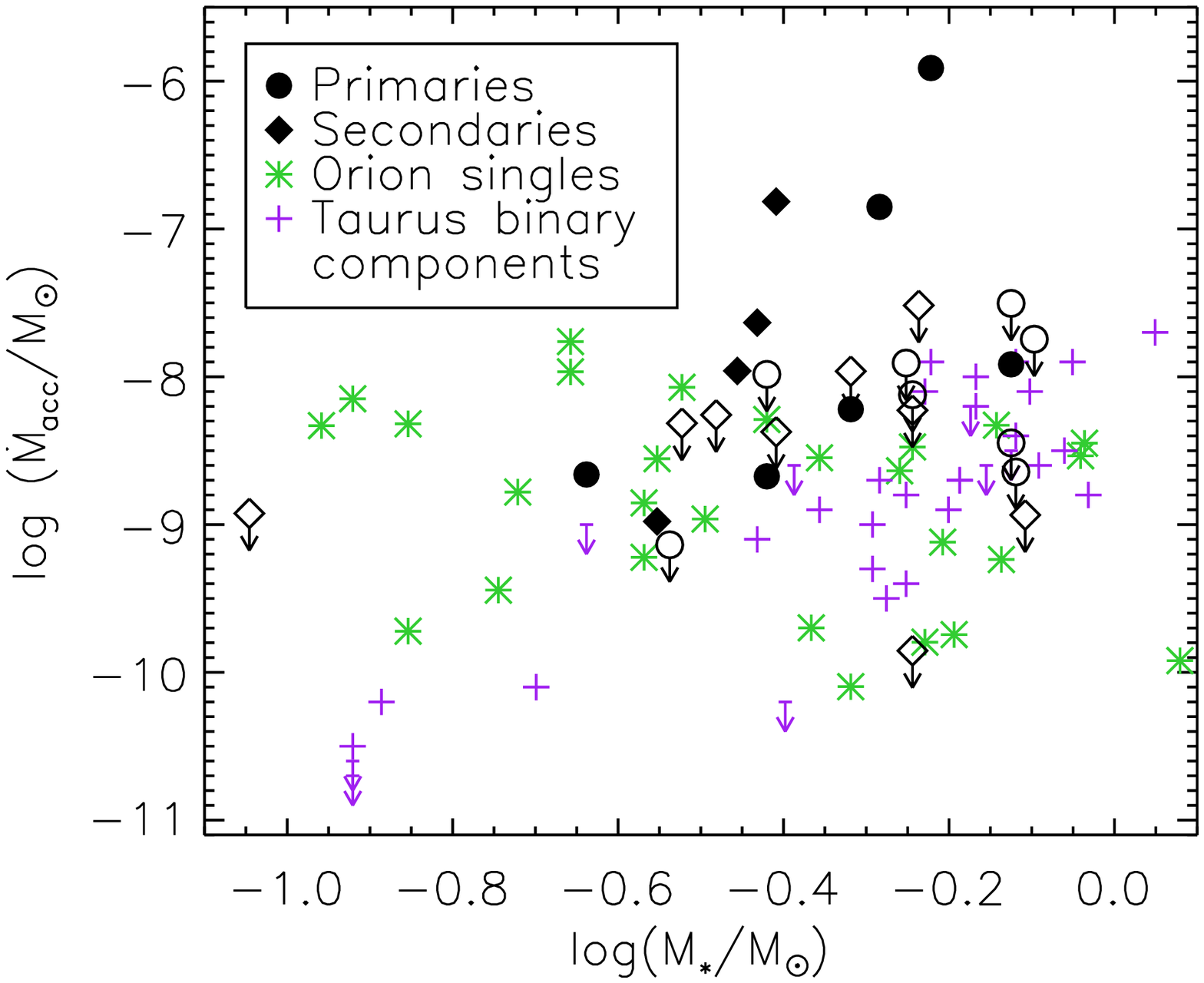}
  \caption{Mass accretion versus stellar mass for all significant emitters of the sample (filled symbols) and upper limits to all other targets with measured $\dot{M}_\mathrm{acc}$ and $M_*$ (open symbols). Primaries are marked with circles, secondaries with diamonds. Asterisks show the mass accretion rates of single stars in Orion \citep{rob04}, whereas plus signs and upper limits are binary components in Taurus \citep{whi01}.
}
  \label{fig:massaccretion}
\end{figure}

Figs.~\ref{fig:accretionluminosityHistogram} and \ref{fig:massaccretion} show a luminosity histogram and component mass-accretion rates as a function of stellar mass, respectively.
In binaries with two accreting components, it is usually the more massive component that has the higher accretion luminosity.
Accordingly, we see a tendency for the subsample of primary components to have slightly higher relative accretion luminosities $L_\mathrm{acc}/L_*$ than the secondaries of our Orion binaries.
The derived accretion luminosities agree very well with single stars in the ONC \citep{rob04,da_10} for $\log(L_\mathrm{acc}/L_\odot)\gtrsim-1.5$, which is the average sensitivity limit of the accretion luminosities derived from our Br$\gamma$ observations.

Similarly, the mass accretion rates of the ONC \mbox{T\,Tauri} binary components as a function of stellar mass are (except for three outliers) comparable to those of single stars in the ONC. Fig.~\ref{fig:massaccretion} shows that Orion single stars occupy almost the same area in the $\log(\dot{M}_\mathrm{acc})$-$\log(M_*)$ diagram, with tendency towards slightly lower accretion rates. This tendency is most likely an observational bias, since \citet{rob04} use $U$-band observations with the Hubble Space Telescope that are more sensitive to lower mass accretion rates than our Br$\gamma$ data.
We also overplot stellar components of Taurus binaries \citep{whi01}, which have similar mass accretion rates.

Comparable mass accretion rates of singles and binaries in the ONC and Taurus\footnote{Compare also singles in Taurus from \citet{muz98}, which are as well at similar accretion rates.} are not self-evident, considering the different disk populations of the ONC and Taurus. Since disk masses in binaries are lower than in single stars of the same star-forming region because of disk truncation \citep{art94} and disks in Orion are less massive than disks in Taurus \citep{man09a}, uniform accretion rates indicate either different disk lifetimes or variable efficiency for the replenishment of an existing disk. The hypothesis of shorter disk lifetimes would corroborate the evidence from Sect.~\ref{sec:synchronizedevolution} that we observe fewer disks around binary components than were measured for single stars.

Three binary components (TCC\,52\,A\&B, JW\,391\,A) have comparably high mass accretion rates to the rest of the sample in Fig.~\ref{fig:massaccretion}. The responsible mechanism is obscure, since the two respective binaries have no remarkable properties in common relative to other targets. Their mass, separation, and distance to $\theta^1$\,Ori\,C are unremarkable. 
While JW\,391A shows no peculiarity in luminosity, the two components of TCC\,52 are found to be the most luminous targets with respect to their mass. The derived very young ages and large radii might indicate an earlier evolutionary stage, i.e.\ lass\,I, which would agree with the higher mass accretion rates of TCC\,52A+B than to older class\,II components \citep{rob06}.
Regardless of their properties, it remains possible that all three components were observed in a temporary state of high activity.

To assess whether disks in binaries need significant replenishment to survive in sufficient quantities, we estimated disk lifetimes $\tau_\mathrm{disk}=M_\mathrm{disk}/\dot{M}_\mathrm{acc}$ from the ratio of disk mass to mass accretion rate and compare it to the age of the star-forming region. To estimate $M_\mathrm{disk}$, we compiled upper limits to the total mass of dusty material around both binary components from the literature of millimeter observations (JW\,519, JW\,681, TCC\,52, TCC\,97, \citealt{man10}; TCC\,52, TCC\,55, \citealt{eis08}).

Only for TCC\,52 do we have available both mass accretion rates and the total mass of the surrounding (disk) material.
The two measurements of total disk mass of TCC\,52 disagree at the 1$\sigma$-level: 0.0288$\pm$0.0029\,$M_\odot$ are derived by \citet{man10} and 0.042$\pm$0.009\,M$_\odot$ by \citet{eis08}. Nevertheless, we now illustrate that an ``order-of-magnitude'' estimate is possible when we assume a total disk mass of $\sim$0.03\,$M_\odot$. 
Disk radii can be estimated from their dynamical truncation radii to be $\sim$0.38 and $\sim$0.3 times the binary separation for primary and secondary, respectively \citep{arm99}, considering our derived binary mass ratio of $q\approx0.65$ and $M_\mathrm{disk}\!\propto\!R_\mathrm{disk}$ \citep[e.g.][]{man10}.
This results in individual disk masses of $M_\mathrm{disk}^\mathrm{prim}\approx0.017\,M_\odot$ and $M_\mathrm{disk}^\mathrm{sec}\approx0.013\,M_\odot$. The derived disk lifetimes for these two target components with the highest mass accretion rates of our sample are $\tau_\mathrm{disk}^\mathrm{prim}\approx1.4\!\times\!10^4$\,yr and $\tau_\mathrm{disk}^\mathrm{sec}\approx8.7\!\times\!10^4$\,yr. Compared to the median age of the ONC binaries of 1\,Myr, derived from Tab.~\ref{tab:componentparameters}, this is very short. These high accretion rates could not have been sustained over the entire early evolution process even if the disk masses were initially ten times more massive than we now observe. If both components were not observed at a younger age than assumed or in a short-lived above-average state of accretion, binary component disks would need substantial replenishment to display the strong accretion activity we detect. 

\subsection{Is the existence of an inner disk linked to planet formation in binaries?}\label{sec:planetformation}
Primordial disks like those detected around the binary components in this sample contain the basic material for the formation of planets. Hence, any peculiarities in the evolution of disks in these systems can leave their footprints on the population of planets in binaries, and it should be instructive to compare the properties of planets and disks around the individual components of binaries.

A recent census identified 40 planets in 35 multiple systems \citep[and references therein]{egg10}, almost all of which orbit the more massive component of the binary \citep{mug09}. An additional eight planets were claimed to reside in circumbinary ({\it P-type}) orbits (PSR\,B1620-26, \citealt{ras94}; HD\,202206, \citealt{cor05}; HW\,Vir, \citealt{lee09}; NN\,Ser, \citealt{beu10}; DP\,Leo, \citealt{qia10}; QS\,Vir, \citealt{qia10a}; HU Aqr, \citealt{qia11}; Kepler-16, \citealt{doy11}). Although some of these planets still need confirmation, we note that all latter candidate hosts are spectroscopic binaries with comparably small separations.
From the thus composed picture of planet occurrence in binaries ($\sim$50 planets in multiples, $\sim$2 around the less massive component, 8 circumbinary planets), one might suspect that (i) planet formation around the less massive components of binaries is suppressed and (ii) that circumbinary planets are rare, but do exist. These observations might either be caused by selection effects, since spectroscopic binaries and fainter secondary stars are less often targeted by spectroscopic surveys, or be the consequence of peculiar disk evolution in binaries.

In an attempt to carefully evaluate systematical errors and biases, \citet{egg10} discovered an underdensity of planets around stars with close, 35--225\,AU stellar companions when compared to single stars with similar properties. This is not seen for wider binaries. The upper limit of 225\,AU noticeably coincides with our 200\,AU transition to synchronized disk evolution. This suggests a common origin of both effects. To test whether shorter lived disks in close binaries can explain the deficiency of planets in binaries of the same separation range, we derived the disk frequency in both subsamples. Binaries with separations $<$200\,AU have a slightly smaller fraction of disk bearing components (34$_{-23}^{+47}$\%) than wide $>$200\,AU pairs (37$_{-27}^{+49}$\%). This difference is, however, not as pronounced as the 1.6--2.1$\sigma$ difference that \citet{egg10} observe for the frequency of planets in the close and wide sample. We concluded that either our sample is not large enough to reveal a significant difference or planet formation in binaries is not strongly correlated with the occurrence of accretion disks.

If the apparent paucity of planets orbiting the less massive components of binaries were not entirely due to selection effects, it could be a consequence of differential disk evolution, which we observe as mixed systems with only one accreting component. If the probability of a star to eventually host a giant extrasolar planet were significantly correlated with the lifetime of its disk, we would expect mainly CW systems, to evolve into circumprimary planetary systems while WC-type binary-disk-systems would preferably evolve into circumsecondary systems. However, we do not observe any significant difference in the appearance of CW systems versus WC (see Sect.~\ref{sec:differentialdiskevolution}). In the same way, we do not see any binaries in which both components are orbited by their individual planets, although the majority of disks evolves synchronizedly. In agreement with this result are the findings of \citet{jen03}, who found that disk masses around the primary are always higher than secondary disk masses in four \mbox{T\,Tauri} binaries, independent of their classification as CC, CW, WC, or WW. Again, a possible explanation could be that the evolution of the inner (accretion) disk is not strongly related to the formation of planets and that other factors dominate the planet formation process.

\section{Summary}\label{sec:summary}
We have presented high-spatial-resolution near-infrared spectroscopic and photometric observations of the individual components of 20 young, low-mass visual binaries in the Orion Nebula Cluster. The sample was complemented with similar observations of six additional targets from \citet[\emph{in prep.}]{cor12}. We have measured the relative positions, $JHK_\mathrm{s}$ photometry, and $K$-band spectra including the accretion-indicating Brackett-$\gamma$ feature in order to derive the projected binary separations as well as the absolute magnitude, spectral type, effective temperature, extinction, veiling, luminosity, and the probability of dust and accretion disks around each binary component. By placing the components into an HR diagram and comparing with pre-main sequence evolutionary tracks, we have estimated the individual age, mass, and radius, as well as the mass accretion rate for each individual component.

Putting the results into context with the star forming environment of the Orion Nebula Cluster and with other young low-mass binary studies, we conclude the following:
\begin{enumerate}
  \item We have found evidence of a slightly lower frequency of circumstellar disks around the individual components of binaries compared to that around single stars of the ONC, in agreement with theory. We have measured a corrected accretion disk fraction of $40_{-9}^{+10}$\% for stars in multiple systems of the ONC, which is lower than the $\sim$50\% accretion disk fraction of single stars in the ONC found by \citet{hil98}. A similar result was found for dust disks, as indicated by NIR excess emission, although with lower significance. As observed for single stars in other clusters, binary components of the ONC have been more often found to contain dust disks than accretion signatures.
  \item The evolution of disks around both components of a binary is correlated for binaries with separations of 200\,AU and below. This was inferred from our inability to detect any mixed pairs of accreting and non-accreting components with separations $<$200\,AU, and that the populations of mixed pairs exhibit significantly larger separations (99.5\% confidence) than pairs of two accreting or two non-accreting components. We have demonstrated that this synchronization is probably not caused by a feeding mechanism involving a circumbinary disk, but possibly instead closer binaries that harbor equal-mass components.
  \item Mixed pairs including an accreting primary and those with an accreting secondary have been observed to be almost equally abundant. In addition to the implication that mixed pairs are common, there is apparently no preference for either the disk of the more or less massive binary component to dissolve first. This points to a \emph{weak} correlation between the binary mass ratio and the presence of a disk around either binary component.
  \item We have found that the mass accretion rates of binary components in the ONC do not differ from the accretion rates of single stars and binary components in Orion and Taurus, respectively. Since disk masses and radii of primordial disks in the ONC are -- on average -- lower, this can potentially lead to shorter lifetimes of disks around binary components, in agreement with our finding of fewer dust and accretion disks than singles of the ONC.
  \item We have measured no strong correlation between the existence of planets around the components of main-sequence binaries and the occurrence of accretion features measured around young binary stars in this paper. Although planets seem to be slightly suppressed in binaries of separations smaller than $\sim$200\,AU \citep[1.6--2.1$\sigma$ significance;][]{egg10}, in wider binaries they are not, and we have not found any equally large differences between the presence of disks in close and wide binary systems.
\end{enumerate}

\begin{acknowledgements}
We thank the referee for a very helpful review leading to a significantly improved paper.
SD would like to thank Thies Heidecke for help with the statistical analysis and Paula S. Teixeira for insightful discussions.
This research has made use of the SIMBAD database, operated at CDS, Strasbourg, France.
In addition, it has used data products from the Two Micron All Sky Survey, which is a joint project of the University of Massachusetts and the Infrared Processing and Analysis Center/California Institute of Technology, funded by the National Aeronautics and Space Administration and the National Science Foundation.
\end{acknowledgements}

\begin{appendices}
\appendix
\section{Statistical evaluation of the disk frequency}\label{sec:app:statistics}
We used a Bayesian approach to derive the probability density function (PDF) of the disk frequency around the individual components of binaries in the ONC. The location of the maximum of the distribution and its width represent the most robust estimates of our disk frequency and its uncertainty, respectively. Input information comes from the measured values of the Br$\gamma$ equivalent width and the continuum noise measurements in each target spectrum.

\subsection{Derivation of the formulae}
We defined the probability density $P(\vartheta|D)$, given the data $D$, of the true disk frequency $\vartheta$ with $0\!\le\!\vartheta\!\le\!1$ to be in the interval [$\vartheta, \vartheta+d\vartheta$], as
\begin{eqnarray}
  P(\vartheta|D) &=& \sum^n_{j=0}\mathrm{prob}\left(\vartheta,k\!=\!j|D\right)\label{eq:A1}\\
                 &=& \sum^n_{j=0}\mathrm{prob}\left(\vartheta|k\!=\!j,D\right)\cdot p_n\left(k\!=\!j|D\right)\quad.\label{eq:A2}
\end{eqnarray} 
While $\mathrm{prob}(\vartheta,k|D)$ describes the PDF as a function of the disk frequency $\vartheta$ and the observed number of disks $k$, the $\mathrm{prob}(\vartheta|k,D)$ in (\ref{eq:A2}) are the probability density functions PDF$_k$, which describe the probability density of finding the true disk frequency to be in the range [$\vartheta, \vartheta+d\vartheta$] when the number of detected disks is fixed at $k$ (see \S\ref{sec:A2}). The latter were weighted by $p_n(k|D)$ in order to account for their relative probability (see \S\ref{sec:A3}). To include the contributions of all possible outcomes (0\dots$n$ disks surrounding $n$ targets), we summed up all partial PDFs to the final PDF $P(\vartheta|D)$.

\subsubsection{The probability density for $k$ disks in $n$ targets}\label{sec:A2}
The PDF $\mathrm{prob}(\vartheta|k,D)$ can be most accurately described as being binomial, since there are only two options ``disk'' or ``no disk''. For instance, starting with Bayes' theorem, we found that
\begin{eqnarray}
  p(\vartheta|k,n) &=& \frac{p\left(k|\vartheta,n\right)p\left(\vartheta|n\right)}{p(k|n)}\\
                   &=& \frac{\left(\!{\begin{array}{*{20}c} n \\ k \end{array}}\!\right)\vartheta^k(1-\vartheta)^{n-k}}{\frac{1}{n+1}}\cdot p(\vartheta|n) \\
                   &=& (n+1)\cdot\left(\!{\begin{array}{*{20}c} n \\ k \end{array}}\!\right)\vartheta^k(1-\vartheta)^{n-k}\cdot \left\{\begin{array}{*{10}l} 1\,\,\mathrm{for\,\,} 0\le\vartheta\le1 \\ 0\,\,\mathrm{else} \end{array}\right.  \quad.
\end{eqnarray}
for the total number of targets $n$. The prior $p(\vartheta|n)$ was assumed to be uniform, since it simply describes the probability of finding a disk fraction anywhere between 0 and 1. In addition, we found that $p(k|n)=1/(n+1)$, since it describes the probability of selecting a solution among $n+1$ possibilities \{$0,\dots,n$\}.

\subsubsection{The weighting factors $p(k|D)$}\label{sec:A3}
We were able to compare the measurement of the equivalent width $W($Br$\gamma)$ to the noise limit at the position of the Br$\gamma$ line (Table~\ref{tab:Aaccretion}). Assuming the noise and the underlying distribution of the Br$\gamma$ measurement to be approximately Gaussian (which can safely be assumed owing to the high photon count rates), we were able to derive the probability $\Pi$ of significant Br$\gamma$ emission and hence a disk by integrating a Gaussian distribution (with median $\mu\!=\!W($Br$\gamma)$ and width $\sigma\!=\!\Delta W($Br$\gamma)$). The lower and upper integration limits were the continuum noise and infinity, respectively. For example, if we measured an equivalent width of 1.5$\pm$0.5\,\AA\ and the continuum noise level is 1.0\,\AA\ (i.e.\ right at the 1$\sigma$\ limit of our measurement), then the derived probability was $\Pi\!=\!0.84$, i.e.\ we assigned a 16\% ($\equiv$1$\sigma$) probability that the system contained no disk.
\begin{table}
  \caption{Accretion}
  \label{tab:Aaccretion}
  \begin{center}
    \begin{tabular}{lcr@{$\,\pm\,$}lc}
      \hline\hline\\[-2ex]
      & 
      &
      \multicolumn{2}{c}{$W_{\mathrm{Br}\gamma}$\tablefootmark{a}} &
      $W_{\mathrm{min}}$\tablefootmark{b}\\ 
      \multicolumn{1}{c}{Name} &
      Component &
      \multicolumn{2}{c}{[\AA]}&
      [\AA]\\[0.5ex] 
      \hline\\[-2ex]
{}[AD95]\,1468  & A &       1.48&0.49 &  1.47  \\            
                & B &       0.68&0.56 &  1.62  \\[0.5ex]     
{}[AD95]\,2380  & A &       1.00&0.19 &  0.65  \\            
                & B &       0.85&1.12 &  3.51  \\[0.5ex]     
JW\,235         & A &       5.13&0.47 &  1.21  \\            
                & B &       2.14&0.54 &  1.59  \\[0.5ex]     
JW\,260         & A &    $-$2.11&0.13 &  0.37  \\            
                & B &    $-$4.05&0.15 &  0.48  \\[0.5ex]     
JW\,553         & A &       0.02&0.14 &  0.44  \\            
                & B &    $-$0.26&0.24 &  0.75  \\[0.5ex]     
JW\,566         & A &       0.63&0.18 &  0.53  \\            
                & B &       0.45&0.23 &  0.70  \\[0.5ex]     
JW\,598         & A &    $-$0.33&0.10 &  0.34  \\            
                & B &       1.10&0.66 &  2.06  \\[0.5ex]     
JW\,648         & A &       0.42&0.17 &  0.49  \\            
                & B &       0.48&0.31 &  0.84  \\[0.5ex]     
JW\,681         & A &    $-$0.38&0.13 &  0.46  \\            
                & B &       0.94&0.21 &  0.68  \\[0.5ex]     
JW\,687         & A &       0.09&0.16 &  0.43  \\            
                & B &    $-$0.05&0.16 &  0.51  \\[0.5ex]     
JW\,876         & A &       0.52&0.18 &  0.54  \\            
                & B &       0.96&0.18 &  0.51  \\[0.5ex]     
JW\,959         & A &    $-$1.48&0.22 &  0.65  \\            
                & B &    $-$0.82&0.25 &  0.76  \\[0.5ex]     
{}[HC2000]\,73  & A &       1.40&0.14 &  0.41  \\            
                & B &    $-$0.10&0.49 &  1.35  \\[0.5ex]     
TCC\,15         & A &    $-$0.79&0.21 &  0.67  \\            
                & B &      16.69&2.73 &  5.44  \\[0.5ex]     
TCC\,52         & A &       1.51&0.10 &  0.28  \\            
                & B &       1.14&0.14 &  0.40  \\[0.5ex]     
TCC\,55         & A &    $-$0.14&0.20 &  0.63  \\            
                & B &    $-$0.24&0.23 &  0.68  \\[0.5ex]     
JW\,63          & A &    $-$0.04&0.34 &  1.02  \\            
                & B &       0.61&0.43 &  1.17  \\[0.5ex]     
JW\,128         & A &       0.53&0.49 &  1.28  \\            
                & B &       0.64&0.49 &  1.28  \\[0.5ex]     
JW\,176         & A &       0.34&0.47 &  1.26  \\            
                & B &       0.22&0.58 &  1.47  \\[0.5ex]     
JW\,391         & A &       7.87&0.26 &  0.69  \\            
                & B &       1.41&0.48 &  1.32  \\[0.5ex]     
JW\,709         & A &    $-$0.16&0.36 &  1.01  \\            
                & B &       0.42&0.35 &  0.97  \\[0.5ex]     
JW\,867         & A &       0.91&0.27 &  0.69  \\            
                & B &       2.37&0.29 &  0.77  \\[0.5ex]     
     \hline     
    \end{tabular}
  \end{center}
\vspace{-3ex}
    \tablefoot{
      \tablefoottext{a} The equivalent width as measured in our spectra. No correction for veiling was applied, since $W_{\mathrm{min}}$ and $W_{\mathrm{Br}\gamma}$ scale the same with veiling.
      \tablefoottext{b} Equivalent width equivalent to 3$\times$ continuum noise level around Br$\gamma$. This is the assumed detection limit for significant Br$\gamma$ emission.
    }
\end{table}

With this knowledge, we were able to derive $p_n(k|D)$, i.e.\ the probabilities of finding $k$ disks in a set of $n$ targets. We defined pairs $\{1-\Pi_i,\Pi_i\}$ of the probabilities of finding ``no disk'' or ``one disk'' around a particular target $i$. The $p_n(k|D)$ were then derived by convolution (zero padded at the edges) of all $\{1-\Pi_i,\Pi_i\}$
\begin{equation}
  p_n(k|D) \in \{1-\Pi_0,\Pi_0\}*\{1-\Pi_1,\Pi_1\}*\dots*\{1-\Pi_n,\Pi_n\}\quad.
\end{equation}
For illustrative purposes, the probabilites of finding 0, 1, or $n$ disks in a sample of $n$ was 
\begin{eqnarray}
  p_n(0|D)&=&\left(1-\Pi_0\right)\left(1-\Pi_1\right)\dots\left(1-\Pi_n\right) = \prod_{j=0}^n\left(1-\Pi_j\right)\quad,\\
  p_n(1|D)&=&\sum_{l=0}^n\left\{\prod_{j\ne l}(1-\Pi_j)\cdot\Pi_l\right\}\quad,\\
  p_n(n|D)&=&\prod_{j=0}^n\Pi_j\quad.
\end{eqnarray}
Other $p_n(k|D)$ could take more complicated forms.

\subsection{Application to the ONC}
\begin{figure}
  \centering
  \includegraphics[width=0.48\textwidth]{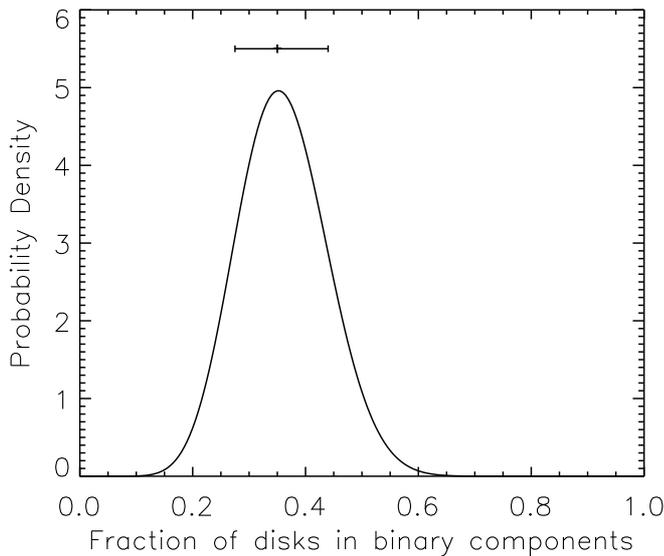}
  \caption{The resulting ONC probability density function as a sum of the individually weighted $p(\vartheta|k,D)$. The resulting best-fit value of the disk fraction of binaries in the ONC (see bar) is $F=0.35^{+0.09}_{-0.08}$.}
  \label{fig:A1}
\end{figure}

For all calculations, we assumed that $n\!=\!42$, which is equal to the number of target components excluding {JW\,260} (see Sect.~\ref{sec:accretiondiskfraction}). The individual $\Pi_{j}$ were calculated from our Br$\gamma$ measurements in Tab.~\ref{tab:Aaccretion}. The final probability density was derived from (\ref{eq:A2}) and is depicted in Fig.~\ref{fig:A1}.

We found a best fit value of $F\!=\!35^{+9}_{-8}$\,\% for the disk frequency in the ONC binaries. The uncertainties of $F$ are defined by the 16\% ($\equiv1\sigma$) wings of the distribution.

\end{appendices}

\bibliographystyle{aa}
\bibliography{ms}

{\Tiny
\longtabL{6}{
\begin{landscape}
\begin{longtable}{lcr@{$\,\pm\,$}lr@{$\,\pm\,$}lr@{$\,\pm\,$}lr@{$\,\pm\,$}lr@{$\,\pm\,$}lr@{$\,\pm\,$}lr@{$\,\pm\,$}lr@{$\,\pm\,$}lrrr@{$\,\pm\,$}lcr@{$\,\pm\,$}lr@{$\,\pm\,$}l}
  \caption{\label{tab:componentparameters}Individual component properties of all targets with detectable photospheric features}\\
    \hline\hline
      & 
      &
      \multicolumn{2}{c}{} &
      \multicolumn{2}{c}{$T_\mathrm{eff}$\tablefootmark{a}} &
      \multicolumn{2}{c}{$A_V$} &
      \multicolumn{2}{c}{$r_K$} &
      \multicolumn{2}{c}{$L_*$} &
      \multicolumn{2}{c}{age\tablefootmark{b}} &
      \multicolumn{2}{c}{$M_*$\tablefootmark{b}} &
      \multicolumn{2}{c}{$R_*$\tablefootmark{b}} &
      \multicolumn{1}{c}{$E_{J-H}$} &
      \multicolumn{1}{c}{$E_{H-K_\mathrm{s}}$} &
      \multicolumn{2}{c}{$W_{\mathrm{Br}\gamma}$\tablefootmark{c}} &
      disk &
      \multicolumn{2}{c}{} &
      \multicolumn{2}{c}{$\dot{M}_\mathrm{acc}$} \\
      \multicolumn{1}{c}{Name} &
      Comp. &
      \multicolumn{2}{c}{SpT} &
      \multicolumn{2}{c}{$[K]$} &
      \multicolumn{2}{c}{[mag]} &
      \multicolumn{2}{r}{$(F_{K_{ex}}/F_{K_{*}})$} &
      \multicolumn{2}{c}{$[L_\odot]$} &
      \multicolumn{2}{c}{$[Myr]$} &
      \multicolumn{2}{c}{$[M_\odot]$} &
      \multicolumn{2}{c}{$[R_\odot]$} &
      \multicolumn{1}{c}{[mag]}&
      \multicolumn{1}{c}{[mag]}&
      \multicolumn{2}{c}{[\AA]}&
      prob.\tablefootmark{d} &
      \multicolumn{2}{c}{log($L_\mathrm{acc}/L_\odot$)} &
      \multicolumn{2}{c}{$[10^{-9}M_\odot\mathrm{yr}^{-1}]$} \\[0.5ex]
    \hline
  \endfirsthead
  \caption{continued.}\\
    \hline\hline
      & 
      &
      \multicolumn{2}{c}{} &
      \multicolumn{2}{c}{$T_\mathrm{eff}$\tablefootmark{a}} &
      \multicolumn{2}{c}{$A_V$} &
      \multicolumn{2}{c}{$r_K$} &
      \multicolumn{2}{c}{$L_*$} &
      \multicolumn{2}{c}{age\tablefootmark{b}} &
      \multicolumn{2}{c}{$M_*$\tablefootmark{b}} &
      \multicolumn{2}{c}{$R_*$\tablefootmark{b}} &
      \multicolumn{1}{c}{$E_{J-H}$} &
      \multicolumn{1}{c}{$E_{H-K_\mathrm{s}}$} &
      \multicolumn{2}{c}{$W_{\mathrm{Br}\gamma}$\tablefootmark{c}} &
      disk &
      \multicolumn{2}{c}{} &
      \multicolumn{2}{c}{$\dot{M}_\mathrm{acc}$} \\
      \multicolumn{1}{c}{Name} &
      Comp. &
      \multicolumn{2}{c}{SpT} &
      \multicolumn{2}{c}{$[K]$} &
      \multicolumn{2}{c}{[mag]} &
      \multicolumn{2}{r}{$(F_{K_{ex}}/F_{K_{*}})$} &
      \multicolumn{2}{c}{$[L_\odot]$} &
      \multicolumn{2}{c}{$[Myr]$} &
      \multicolumn{2}{c}{$[M_\odot]$} &
      \multicolumn{2}{c}{$[R_\odot]$} &
      \multicolumn{1}{c}{[mag]}&
      \multicolumn{1}{c}{[mag]}&
      \multicolumn{2}{c}{[\AA]}&
      prob.\tablefootmark{d} &
      \multicolumn{2}{c}{log($L_\mathrm{acc}/L_\odot$)} &
      \multicolumn{2}{c}{$[10^{-9}M_\odot\mathrm{yr}^{-1}]$} \\[0.5ex]
    \hline
  \endhead
  \hline
  \endfoot
{}[AD95]\,1468          & A &    M4  &     1             &    3270&150                &    3.7  &     1.1                  &    0.00   &  0.29          &       0.11  &  0.03        &           4.0 &$_{1.4}^{4.0}$  &        0.23 &$_{0.05}^{0.05}$&         1.04 & 0.05         &    0.43   &   0.47  &   1.48&0.50                    &  0.50    & $-$1.92&0.21                &   2.2 &1.44                 \\                
                        & B &\multicolumn{2}{c}{$\cdots$}&\multicolumn{2}{c}{$\cdots$}&    0.5  &     1.3                  &\multicolumn{2}{c}{$\cdots$}&\multicolumn{2}{c}{$\cdots$}&\multicolumn{2}{c}{$\cdots$}    &\multicolumn{2}{c}{$\cdots$}  &\multicolumn{2}{c}{$\cdots$} &$\cdots$&$\cdots$    & \multicolumn{2}{c}{($<$1.62)} &  0.05     &\multicolumn{2}{c}{$\cdots$ }&\multicolumn{2}{c}{$\cdots$ }\\[0.5ex]         
{}[AD95]\,2380          & A &\multicolumn{2}{c}{$\cdots$}&\multicolumn{2}{c}{$\cdots$}& \multicolumn{2}{c}{$\cdots$}       &\multicolumn{2}{c}{$\cdots$}&\multicolumn{2}{c}{$\cdots$}&\multicolumn{2}{c}{$\cdots$}    &\multicolumn{2}{c}{$\cdots$}  &\multicolumn{2}{c}{$\cdots$} &$\cdots$&$\cdots$    & (1.00&0.19)                   &  0.97     &\multicolumn{2}{c}{$\cdots$ }&\multicolumn{2}{c}{$\cdots$ }\\                
                        & B &\multicolumn{2}{c}{$\cdots$}&\multicolumn{2}{c}{$\cdots$}& \multicolumn{2}{c}{$\cdots$}       &\multicolumn{2}{c}{$\cdots$}&\multicolumn{2}{c}{$\cdots$}&\multicolumn{2}{c}{$\cdots$}    &\multicolumn{2}{c}{$\cdots$}  &\multicolumn{2}{c}{$\cdots$} &$\cdots$&$\cdots$    & \multicolumn{2}{c}{($<$3.51)} &  0.01     &\multicolumn{2}{c}{$\cdots$} &\multicolumn{2}{c}{$\cdots$} \\[0.5ex]         
JW\,235                 & A &\multicolumn{2}{c}{$\cdots$}&\multicolumn{2}{c}{$\cdots$}&    0.0  &     1.0                  &\multicolumn{2}{c}{$\cdots$}&\multicolumn{2}{c}{$\cdots$}&\multicolumn{2}{c}{$\cdots$}    &\multicolumn{2}{c}{$\cdots$}  &\multicolumn{2}{c}{$\cdots$} & $\cdots$  & $\cdots$&  (5.13&0.47)                   &  1.00    &\multicolumn{2}{c}{$\cdots$ }&\multicolumn{2}{c}{$\cdots$ }\\                
                        & B &    M3  &     3             &    3420&440                &    5.7  &     0.9                  &    0.73   &  0.65          &       0.62  &  0.16        &           1.0 &$_{0.9}^{2.0}$  &        0.35 &$_{0.17}^{0.23}$&         2.24 & 0.29         & $-$0.04   &$-$0.26  &   3.67&0.93                    &  0.84    & $-$1.37&0.18                &     11&$_{   8}^{   9}$     \\[0.5ex]         
JW\,260                 & A &    G0  &     1             &    6030&170                &    0.0  &     1.1                  &    0.20   &  0.04          &      18.0   &  5.5         &           3.5 &0.5             &        2.5  &$_{0.3}^{0.2}$  &         3.89 & 0.11         &    0.29   &   0.19  &$-$2.52&0.15                    &  0.00    &\multicolumn{2}{c}{$\cdots$} &\multicolumn{2}{c}{$\cdots$} \\                
                        & B &    F7  &     3             &    6320&210                &    0.0  &     1.1                  &    0.00   &  0.04          &      12.1   &  3.7         &           4.0 &0.5             &        2.3  &$_{0.2}^{0.2}$  &         2.91 & 0.10         &    0.44   &   0.43  &$-$4.05&0.15                    &  0.00    &\multicolumn{2}{c}{$\cdots$} &\multicolumn{2}{c}{$\cdots$} \\[0.5ex]         
JW\,553                 & A &    K7  & $^{1}_2$          &    4060&250                &    3.0  &     0.1                  &    0.22   &  0.06          &       5.2   &  1.5         &           0.35&$_{0.32}^{0.35}$&        0.8  &$_{0.2}^{0.3}$  &         4.63 & 0.28         & $-$0.01   &$-$0.03  &  \multicolumn{2}{c}{$<$0.54}   &  0.00    &\multicolumn{2}{c}{$<-1.11$} &\multicolumn{2}{c}{$<18  $}  \\                
                        & B &    M1  &     3             &    3710&440                &    1.4  &     1.7                  &    1.72   &  0.57          &       0.16  &  0.02        &           8   &$_{5}^{32}$     &        0.57 &$_{0.3}^{0.13}$ &         0.97 & 0.12         & $-$0.14   &   0.26  &  \multicolumn{2}{c}{$<$2.04}   &  0.00    &\multicolumn{2}{c}{$<-2.68$} &\multicolumn{2}{c}{$<0.14$}  \\[0.5ex]         
JW\,566                 & A &    K7  &    1.5            &    4060&280                &    4.4  &     1.1                  &    0.02   &  0.05          &       1.3   &  0.2         &           1.2 &$_{0.5}^{1.8}$  &        0.75 &$_{0.23}^{0.35}$&         2.33 & 0.16         &    0.49   &   0.78  &   0.64&0.19                    &  0.70    & $-$1.01&0.05                &    12 &$_{ 4  }^{ 6  }$     \\                
                        & B &    M1.5&     2             &    3705&290                &    0.0  &     0.3                  &    0.00   &  0.05          &       1.2   &  0.1         &           0.9&$_{0.7}^{0.5}$   &        0.48 &$_{0.13}^{0.2}$ &         2.61 & 0.20         &    0.21   &   0.29  &  \multicolumn{2}{c}{$<$0.70}   &  0.13    &\multicolumn{2}{c}{$<-1.30$} &\multicolumn{2}{c}{$<11  $}  \\[0.5ex]         
JW\,598\tablefootmark{e}& A &    K5  &     2             &    4350&380                &    4.4  &     0.4                  &\multicolumn{2}{c}{$\cdots$}&\multicolumn{2}{c}{$\cdots$}& \multicolumn{2}{c}{$\cdots$}   & \multicolumn{2}{c}{$\cdots$} &\multicolumn{2}{c}{$\cdots$} &         &           &  \multicolumn{2}{c}{($<$0.34)} &  0.00    &\multicolumn{2}{c}{$\cdots$} &\multicolumn{2}{c}{$\cdots$} \\                
                        & B &    M2  &     4             &    3560&570                &    5.0  &     0.4                  &    0.01   &  0.44          &\multicolumn{2}{c}{$\cdots$}& \multicolumn{2}{c}{$\cdots$}   & \multicolumn{2}{c}{$\cdots$} & \multicolumn{2}{c}{$\cdots$}& $\cdots$  & $\cdots$&  \multicolumn{2}{c}{$<$2.08}   &  0.07    &\multicolumn{2}{c}{$\cdots$ }&\multicolumn{2}{c}{$\cdots$ }\\[0.5ex]         
JW\,648                 & A &    M0  &     1             &    3850&170                &    2.3  &     0.1                  &    0.06   &  0.05          &       1.7   &  0.1         &           0.7 &0.2             &        0.57 &$_{0.12}^{0.13}$&         2.90 & 0.13         &    0.16   &   0.30  &  \multicolumn{2}{c}{$<$0.52}   &  0.33    &\multicolumn{2}{c}{$<-1.43$} &\multicolumn{2}{c}{$<7.6 $}  \\                
                        & B &    M3.5&     2             &    3340&290                &    2.3  &     0.3                  &    0.00   &  0.14          &       0.49  &  0.04        &           1.0 &0.8             &        0.30 &$_{0.10}^{0.11}$&         2.09 & 0.18         &    0.21   &   0.14  &  \multicolumn{2}{c}{$<$0.84}   &  0.12    &\multicolumn{2}{c}{$<-1.76$} &\multicolumn{2}{c}{$<4.9 $}  \\[0.5ex]         
JW\,681                 & A &    M1  &    1.5            &    3710&220                &    8.1  &     0.6\tablefootmark{f} &    0.09   &  0.06          &\multicolumn{2}{c}{$\cdots$}& \multicolumn{2}{c}{$\cdots$}   & \multicolumn{2}{c}{$\cdots$}& \multicolumn{2}{c}{$\cdots$}& $\cdots$  & $\cdots$&  \multicolumn{2}{c}{$<$0.50}   &  0.00    &\multicolumn{2}{c}{$\cdots$ }&\multicolumn{2}{c}{$\cdots$ }\\                                             
                        & B &    M3.5&     4             &    3340&550                &    7.6  &     1.2\tablefootmark{f} &    0.26   &  0.15          &\multicolumn{2}{c}{$\cdots$}& \multicolumn{2}{c}{$\cdots$}   & \multicolumn{2}{c}{$\cdots$}& \multicolumn{2}{c}{$\cdots$}& $\cdots$  & $\cdots$&   1.18&0.26                    &  0.90    &\multicolumn{2}{c}{$\cdots$ }&\multicolumn{2}{c}{$\cdots$ }\\[0.5ex]                                      
JW\,687                 & A &    M2.5&$^{0.5}_1$         &    3490&150                &    6.1  &     0.1                  &    0.43   &  0.10          &       1.7   &  0.1         &           0.10&0.05            &        0.38 &$_{0.06}^{0.05}$&         3.58 & 0.15         &    0.03   &$-$0.06  &  \multicolumn{2}{c}{$<$0.61}   &  0.02    &\multicolumn{2}{c}{$<-1.56$} &\multicolumn{2}{c}{$<10  $}  \\                
                        & B &    M2  &     1             &    3560&150                &    7.4  &     0.4                  &    1.62   &  0.27          &       0.54  &  0.06        &           1.3 &$_{0.4}^{0.5}$  &        0.39 &$_{0.06}^{0.05}$&         1.93 & 0.08         &    0.34   &   0.49  &  \multicolumn{2}{c}{$<$1.34}   &  0.00    &\multicolumn{2}{c}{$<-1.67$} &\multicolumn{2}{c}{$<4.2 $}  \\[0.5ex]         
JW\,876                 & A &    M0.5&     1             &    3790&210                &    2.3  &     0.6\tablefootmark{f} &    0.02   &  0.06          &\multicolumn{2}{c}{$\cdots$}& \multicolumn{2}{c}{$\cdots$}   & \multicolumn{2}{c}{$\cdots$} & \multicolumn{2}{c}{$\cdots$}& $\cdots$  & $\cdots$&  \multicolumn{2}{c}{$<$0.55}   &  0.47    &\multicolumn{2}{c}{$<-0.87$} &\multicolumn{2}{c}{$\cdots$ }\\                
                        & B &    M1.5&     3             &    3630&440                &    0.0  &     1.0\tablefootmark{f} &    0.38   &  0.13          &\multicolumn{2}{c}{$\cdots$}& \multicolumn{2}{c}{$\cdots$}   & \multicolumn{2}{c}{$\cdots$} & \multicolumn{2}{c}{$\cdots$}& $\cdots$  & $\cdots$&   1.30&0.24                    &  0.99    & $-$0.97&0.04                &\multicolumn{2}{c}{$\cdots$ }\\[0.5ex]         
JW\,959                 & A &    K3  &     3             &    4730&450                &    3.0  &     0.8\tablefootmark{f} &    0.27   &  0.10          &\multicolumn{2}{c}{$\cdots$}& \multicolumn{2}{c}{$\cdots$}   & \multicolumn{2}{c}{$\cdots$} & \multicolumn{2}{c}{$\cdots$}& $\cdots$  & $\cdots$&$-$1.86&0.27                    &  0.00    &\multicolumn{2}{c}{$\cdots$} &\multicolumn{2}{c}{$\cdots$} \\                
                        & B &    K3  &     2             &    4730&370                &    5.6  &     0.7\tablefootmark{f} &    0.00   &  0.08          &\multicolumn{2}{c}{$\cdots$}& \multicolumn{2}{c}{$\cdots$}   & \multicolumn{2}{c}{$\cdots$} & \multicolumn{2}{c}{$\cdots$}& $\cdots$  & $\cdots$&$-$0.81&0.25                    &  0.00    &\multicolumn{2}{c}{$\cdots$} &\multicolumn{2}{c}{$\cdots$} \\[0.5ex]         
{}[HC2000]\,73          & A &    M2  &     1             &    3560&150                &    1.1  &     0.5                  &    1.63   &  0.28          &       0.21  &  0.03        &           3.5 &$_{1.0}^{3.0}$  &        0.38 &$_{0.06}^{0.10}$&         1.21 & 0.05         &    0.34   &   0.51  &   3.68&0.37                    &  1.00    & $-$1.78&0.16                &   2.1 &$_{  1.0 }^{  1.1 }$ \\                
                        & B &    M7  &     1             &    2880&140                &    0.0  &     0.5                  &    0.01   &  0.24          &       0.08  &  0.01        &           0.6 &$_{0.4}^{3.0}$  &        0.09 &$_{0.05}^{0.05}$&         1.17 & 0.06         & $-$0.22   &$-$0.02  &  \multicolumn{2}{c}{$<$1.36}   &  0.00    &\multicolumn{2}{c}{$<-2.64$} &\multicolumn{2}{c}{$<1.2 $}  \\[0.5ex]         
TCC\,15\tablefootmark{e}& A &    K4  &     2             &    4590&290                &   10.6  &     0.4                  &\multicolumn{2}{c}{$\cdots$}&         2.7 &  0.3         &           1.8 &$_{1.0}^{3.2}$  &        1.5  &$_{0.5}^{0.2}$  &         2.63 & 0.17         &    0.09   &   0.06  &($-$0.79&0.21)                  &  0.00    &\multicolumn{2}{c}{$\cdots$} &\multicolumn{2}{c}{$\cdots$} \\        [0.5ex] 
                        & B &\multicolumn{2}{c}{$\cdots$}&\multicolumn{2}{c}{$\cdots$}&    3.0  &     2.8                  &\multicolumn{2}{c}{$\cdots$}&\multicolumn{2}{c}{$\cdots$}&\multicolumn{2}{c}{$\cdots$}    &\multicolumn{2}{c}{$\cdots$}  &\multicolumn{2}{c}{$\cdots$} &$\cdots$&$\cdots$    & (16.69&2.73)                   &  1.00    & \multicolumn{2}{c}{$\cdots$}&\multicolumn{2}{c}{$\cdots$ }\\[0.5ex]         
TCC\,52                 & A &    M0.5&     2             &    3790&300                &    0.7  &     0.3                  &    4.42   &  0.36          &      10.4   &  0.9         &           0.01&$_{0.01}^{0.20}$&        0.6  &$_{0.2}^{0.3}$  &         7.50 & 0.59         &    0.36   &   0.59  &   8.00&0.56                    &  1.00    &    0.39&0.20                &   1230&$_{830}^{950}$       \\                
                        & B &    M2  &     1             &    3560&150                &    2.0  &     0.4                  &    0.80   &  0.12          &       3.1   &  0.4         &           0.05&$_{0.04}^{0.15}$&        0.39 &$_{0.05}^{0.08}$&         4.61 & 0.19         &    0.28   &   0.40  &   2.04&0.25                    &  1.00    & $-$0.49&0.07                &    153&$_{ 34}^{ 42}$       \\[0.5ex]         
TCC\,55                 & A &    M3  &     1             &    3420&150                &    3.0  &     0.7                  &    0.14   &  0.10          &       0.09  &  0.02        &           6   &$_{2}^{6}$      &        0.29 &$_{0.06}^{0.07}$&         0.87 & 0.04         &    0.86   &   1.23  &  \multicolumn{2}{c}{$<$0.72}   &  0.00    &\multicolumn{2}{c}{$<-2.19$} &\multicolumn{2}{c}{$<0.7 $}  \\                
                        & B &\multicolumn{2}{c}{$\cdots$}&\multicolumn{2}{c}{$\cdots$}&    0.0  &     0.7                  &\multicolumn{2}{c}{$\cdots$}&\multicolumn{2}{c}{$\cdots$}&\multicolumn{2}{c}{$\cdots$}    &\multicolumn{2}{c}{$\cdots$}  &\multicolumn{2}{c}{$\cdots$} &$\cdots$&$\cdots$    & \multicolumn{2}{c}{($<$0.68)} &  0.00     &\multicolumn{2}{c}{$\cdots$} &\multicolumn{2}{c}{$\cdots$} \\[0.5ex]         
JW\,63                  & A &    K7  &     2             &    4060&320                &    2.0  &     0.3                  &    0.00   &  0.05          &       1.3   &  0.1         &           1.2 &$_{0.5}^{1.8}$  &        0.75 &$_{0.25}^{0.4}$ &         2.32 & 0.18         & $-$0.06   &$-$0.10  &  \multicolumn{2}{c}{$<$1.02}   &  0.00    &\multicolumn{2}{c}{$<-1.54$} &\multicolumn{2}{c}{$<3.6 $}  \\                
                        & B &    M3  &     2             &    3420&290                &    3.2  &     0.3                  &    0.00   &  0.21          &       0.84  &  0.07        &           0.7 &$_{0.6}^{0.5}$  &        0.33 &$_{0.10}^{0.15}$&         2.62 & 0.22         & $-$0.16   &$-$0.44  &  \multicolumn{2}{c}{$<$1.17}   &  0.09    &\multicolumn{2}{c}{$<-1.76$} &\multicolumn{2}{c}{$<5.5 $}  \\[0.5ex]         
JW\,128                 & A &    M0  &     2             &    3850&390                &    2.2  &     0.2                  &    0.00   &  0.11          &       1.6   &  0.1         &           0.7 &$_{0.6}^{1.0}$  &        0.56 &$_{0.19}^{0.39}$&         2.81 & 0.28         & $-$0.05   &$-$0.05  &  \multicolumn{2}{c}{$<$1.28}   &  0.06    &\multicolumn{2}{c}{$<-1.21$} &\multicolumn{2}{c}{$<12  $}  \\                
                        & B &    M0  &     2             &    3850&390                &    0.8  &     0.2                  &    0.00   &  0.13          &       0.89  &  0.06        &           1.2 &$_{0.5}^{2.5}$  &        0.57 &$_{0.19}^{0.40}$&         2.12 & 0.21         &    0.07   &   0.14  &  \multicolumn{2}{c}{$<$1.26}   &  0.09    &\multicolumn{2}{c}{$<-1.40$} &\multicolumn{2}{c}{$<5.9 $}  \\[0.5ex]         
JW\,176                 & A &    K7  &     2             &    4060&320                &    3.2  &     0.2                  &    0.02   &  0.16          &       3.5   &  0.2         &           0.45&$_{0.40}^{0.45}$&        0.75 &$_{0.25}^{0.40}$&         3.80 & 0.30         & $-$0.04   &$-$0.08  &  \multicolumn{2}{c}{$<$1.50}   &  0.02    &\multicolumn{2}{c}{$<-0.81$} &\multicolumn{2}{c}{$<31  $}  \\                
                        & B &    M0  &     2             &    3850&390                &    4.0  &     0.2                  &    0.00   &  0.19          &       2.8   &  0.2         &           0.45&0.40            &        0.58 &$_{0.21}^{0.40}$&         3.75 & 0.38         & $-$0.18   &$-$0.26  &  \multicolumn{2}{c}{$<$1.47}   &  0.02    &\multicolumn{2}{c}{$<-0.93$} &\multicolumn{2}{c}{$<30  $}  \\[0.5ex]         
JW\,391                 & A &    M0.5&     3             &    3790&440                &    1.2  &     0.2                  &    0.71   &  0.31          &       1.1   &  0.1         &           0.9 &$_{0.8}^{1.6}$  &        0.52 &$_{0.10}^{0.40}$&         2.42 & 0.28         &    0.21   &   0.35  &  12.91&0.45                    &  1.00    & $-$0.12&0.11                &    140&$_{ 50}^{120}$       \\                
                        & B &    M3.5&     2             &    3340&290                &    0.1  &     0.2                  &    0.00   &  0.22          &       0.15  &  0.01        &           3.5 &$_{1.5}^{3.5}$  &        0.28 &$_{0.11}^{0.12}$&         1.16 & 0.10         &    0.20   &   0.11  &   1.41&0.48                    &  0.58    & $-$2.20&0.22                &   1.1 &$_{0.8 }^{0.8 }$     \\[0.5ex]         
JW\,709                 & A &    K7  &     2             &    4060&320                &    1.7  &     0.2                  &    0.00   &  0.24          &       0.94  &  0.06        &           1.8 &$_{0.9}^{3.2}$  &        0.76 &$_{0.25}^{0.40}$&         1.97 & 0.16         & $-$0.00   &$-$0.01  &  \multicolumn{2}{c}{$<$1.01}   &  0.00    &\multicolumn{2}{c}{$<-1.66$} &\multicolumn{2}{c}{$<2.3 $}  \\                
                        & B &    K7  &     1             &    4060&250                &    1.6  &     0.2                  &    0.00   &  0.11          &       0.69  &  0.05        &           3.0 &$_{1.6}^{4.0}$  &        0.78 &$_{0.23}^{0.40}$&         1.68 & 0.10         & $-$0.01   &$-$0.02  &  \multicolumn{2}{c}{$<$0.97}   &  0.06    &\multicolumn{2}{c}{$<-1.87$} &\multicolumn{2}{c}{$<1.2 $}  \\[0.5ex]         
JW\,867                 & A &    M1  &     1             &    3710&150                &    0.1  &     0.3                  &    0.13   &  0.12          &       0.89  &  0.08        &           1.0 &$_{0.2}^{0.4}$  &        0.48 &$_{0.10}^{0.10}$&         2.29 & 0.09         &    0.16   &   0.24  &   1.03&0.31                    &  0.80    & $-$1.50&0.12                &   6.0 &2.3                  \\                
                        & B &    M2.5&     1             &    3490&150                &    0.0  &     0.3                  &    0.02   &  0.11          &       0.77  &  0.07        &           0.9 &$_{0.6}^{0.2}$  &        0.37 &$_{0.06}^{0.06}$&         2.41 & 0.10         &    0.14   &   0.19  &   2.41&0.29                    &  1.00    & $-$1.05&0.05                &   23  &4.8                  \\[0.5ex]         
     \hline                                                                              
\end{longtable}                                                                          
    \vspace{-4ex}\begin{minipage}{1.29\textwidth}\tablefoot{                             
      \tablefoottext{a}Effective temperatures were converted with the temperature scale for pre-main sequence stars given in \citet{luh03} for spectral types later than M0. The effective temperatures of earlier spectral types are from \citet{sch82}. Uncertainties are from $T_\mathrm{eff}$ at the SpT uncertainty limits.
      \tablefoottext{b}Results from comparing $T_\mathrm{eff}$ and $L_*$ with the models of \citet{sie00}. 
      \tablefoottext{c}Positive values indicate emission lines, negative values indicate Br$\gamma$ in absorption. Upper limits refer to the minimum equivalent widths of an emission line at $\lambda_\mathrm{Br\gamma}$ that can be detected with 3$\sigma$ significance. Whenever possible, equivalent widths were extracted from the $r_K$-corrected spectra. Values from targets without $r_K $are in parantheses. These can be regarded as lower limits to the equivalent width of the Br$\gamma$ line and no accretion luminosities or mass accretion rates were derived.
      \tablefoottext{d}The disk probability is the result of the integration described in \S\ref{sec:A3}. This does not exclude the possibility of 'filled-up' absorption lines where emission and absorption cancel each other.
      \tablefoottext{e}The spectral features allow us to assign a spectral type, but the spectral fitting routine was unable to match the overall continuum shape of this target, in particular for a strong increase at $\lambda>24000$\,\AA.
      \tablefoottext{f}These extinction values were derived from our spectral fitting instead of photometry, since our photometric information was incomplete.
    }\end{minipage}                                                                   
\end{landscape}                                                                       
}
}

\end{document}